\documentclass[12pt]{article}
\pdfoutput=1

\usepackage{graphicx}
\usepackage[pdftex,bookmarks=true]{hyperref}
 \usepackage{hyperref}
\usepackage{ amsmath, amsthm, amssymb, amsfonts}
\begin{document}
\vspace*{-15ex}
\textsc{\hfill  Sep 15, 2012}
\begin{center}{\large \bf Two Speed TASEP with Step Initial Condition}\end{center}

\begin{center}{\large \bf Josh Jen Keng OYoung}\\
{\it Department of Mathematics\\
University of California\\
Davis, CA 95616, USA\\
email: oyounggo@math.ucdavis.edu}\end{center}

\begin{abstract}
In this paper, we consider zero range process with an initial condition which is equivalent to step initial condition in total asymmetric simple exclusion process (TASEP) as described in a paper by R\'akos, A. and Sch\"utz \cite{Ra3}by using techniques developed by Borodin, Ferrari, and Sasamoto \cite{Bo2}.
The solution for the transition probability of total asymmetric simple exclusion process for particles with different hopping rates was first worked out by Sch\"utz and Rak\"os \cite{Ra} for the case when $p=1$ or $q=1$. The formula was later applied to analyze two speed TASEP $\cite{Bo2}$ with alternating initial condition. Here we will investigate the two speed TASEP case with step initial condition.

\end{abstract}

\section{Introduction}
\label{sec:LabelForIntroduction}

The asymmetric simple exclusion process (ASEP) was first introduced in 1970 by Frank Spitzer \cite{Sp}. It has a wide range of applications \cite{Li}\cite{SU} and it is now known to belong to the Kardar-Parisi-Zhang (KPZ) universality class \cite{KPZ}.

ASEP is a model for particles interacting on a lattice. Each particle is equipped with an alarm clock
with exponential distribution of hopping rate $v_i$: The particle jumps when the clock rings. It has a
probability $p$ jumps to the right and $q = 1-p$ jumps to the left provided the neighboring site is empty.
When $p = 1$, we call it total asymmetric simple exclusion process (TASEP). Kurt Johansson\cite{Jo} was the first
to show that when $N$ particles with equal hopping rates are initially assigned in the negative region
called step initial condition, meaning the position of the $i$th particle $x_i(t) = -i + 1$ when $t = 0$, the
following formula connecting TASEP with random matrix theory holds:
\begin{equation}
P(x_N(t)\geq t)=\int_0^t...\int^t_0\prod_{i=1}^Nt_i^{x+N-1}e^{-t_i}\prod_{i<j}(t_i-t_j)^2  dt_1\cdots dt_N.
\end{equation}

Here, the right hand side of (1.1.1) with the Laguerre weight function ($w(t_i)=t_i^{x+N-1}e^{-t_i}$), is equal to the probability that the largest eigenvalue of a random matrix $AA$* is $\leq t$ in Laguerre ensemble where $A$ is a $N\times(x-1+2N)$ matrix of complex Gaussian random variables with mean zero and variance $1/2$.

Later on Gunter M. Sch\"utz and A. R\'akos \cite{Ra} and Tomohiro Sasamoto and Taro Nago \cite{Sa} also derived the same formula above using Bethe Ansatz technique. Recently Craig Tracy and Harold Widom \cite{Tr} have also shown that Bethe Ansatz technique can be applied to ASEP process. They also extended Johansson result to ASEP process.

In this paper, we will use the Fredholm determent formula for $P(x_1(t),...,x_N(t);y_1,....,y_N)$ from a paper by Borodin and Ferrari $\cite{Bo0}$. Then we will work out the bi-orthornormal polynomials for the correlation kernel of $P(x_1(t),...,x_N(t);y_1,....,y_N)$. We will see that when the particles starts with step initial condition while the first $M$ leading particles moving with rate $\alpha\in(0,1)$, slower than the rest of the particles which have unit speed, our process diagram, as shown in section 2, has a frozen region\footnote{Frozen region means particles sitting in this region are standing still since their neighboring sites are all occupied.} and it is different from the case with periodic initial condition\cite{Bo2}. Also with step initial condition, we won't have a so called shock line as the case in the paper by Borodin, Ferrari, and Sasamoto \cite{Bo2}. We will also discuss how we can apply the formula in section 3 to study zero range process as certain parameters goes to infinite as what R\'akos, A. and Sch\"utz discussed in their paper \cite{Ra3}.

%
%
%

\section{Main Results}
\label{sec:LabelForMain Results}
We will consider TASEP With step initial condition $x_i(0)=y_i=M-i$ and letting the first $M$ particles have rate
$\alpha$ and the rest of particles with unit speed(see figure below for an example with $M=3$).

\includegraphics[width=5in,height=1in]{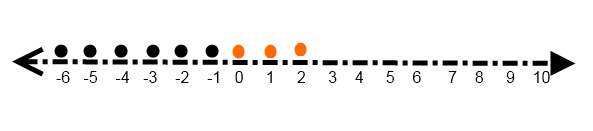}\


In this section, we will study the asymptotics of $P(x_n(t)\geq x)$. The main results to be proved in this section is the following process diagram which gives the relationship between $n/t$ and $\alpha.$ Here $n$ and $t$ are both parameters in $x_n(t)$ and $\alpha$ is the speed of the right most particle. In the diagram $A_2$ is the Airy$_2$ process, $A_{DBM}$ is the Dyson's Brownian motion process and the curved line $(1-\alpha)^2$ in the diagram represents the $A_{DBM\to 2}$ process. The definition of these process will be given in appendix. It should be noted here our diagram has a frozen region and it is different from the case with periodic initial condition\cite{Bo2}.\

\includegraphics[width=6in,height=3in]{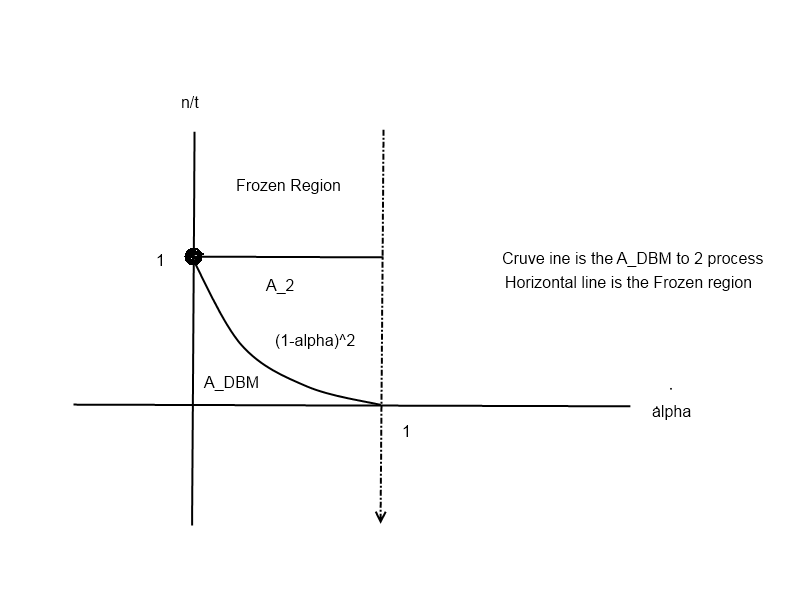}

A by product of our result is to use our results to study zero range process which will be explained in the following paragraphs.

Zero range process is a process similar to TASEP but here our particles do not exclude each other and all particles' hopping rates depend on their which site they are sitting on. In our case we will assume every particle has rate $1$ hopping to the left from site 1 to site $L-1$ and there is an injection of particles to site $L-1$ from the right with rate $\alpha.$ We will call this kind of zero range process totally asymmetric zero range process. It is easy to see the current distribution across the ${(L-1)}^{\text{th}}$ bond of our total zero range process (TAZRP), $p_{L-1}(x,t)$, is equal to the $(L)$-particle TASEP case $\mathbb{P}(x_{L+1}(t)\geq x).$ When we map the $L$ sites zero range process to $(L+1)$ particles TASEP, we have $v_i=1$ for all $i$ but $v_{L}=\alpha.$ This approach of using Bethe ansatz solution to study zero range process was also used by Rakos and Schutz\cite{Ra3}. In this paper, we take the limit $L=\nu t$ then we have

\begin{equation}\lim_{t\to\infty}p_{L(t)-1}(t)=\lim_{t\to\infty}\mathbb{P}(x_{[\nu t]}(t)\leq s )  .\end{equation}

The picture of mapping zero range process to TASEP is given below.\

\

\includegraphics[width=5in,height=3.7in]{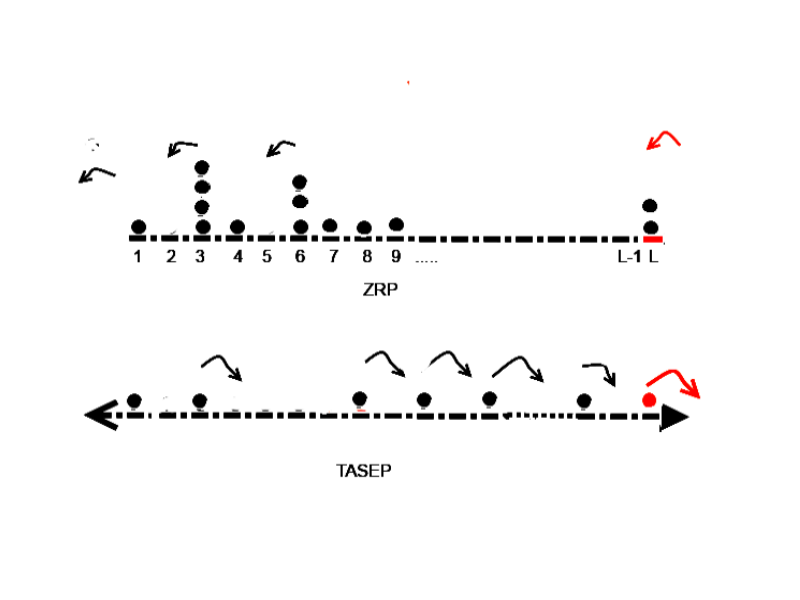}

Now we will discuss our model and our main results.
If particles starts from $\mathbb{Z_{-}}$ then the particles density $\mathbb{P}$(there is a particle at $x$ at time $t$)$:=u(x,t)$ is given as below \cite{KT}

\begin{equation}
u(x,t)=
\left\{\begin{array}{ll}
1,& x<-t,\\
1/2-x/t/2,&x\in [-t,(2\alpha-1)t],\\
1-\alpha,& x>(2\alpha-1)t.
\end{array}\right.
\end{equation}

while the initial condition is given as the following

\begin{equation}
u(x,0):=
\left\{\begin{array}{ll}
1,& x<0,\\

1-\alpha,& x\geq 0.
\end{array}\right.
\end{equation}

\


\includegraphics[width=6in,height=3in]{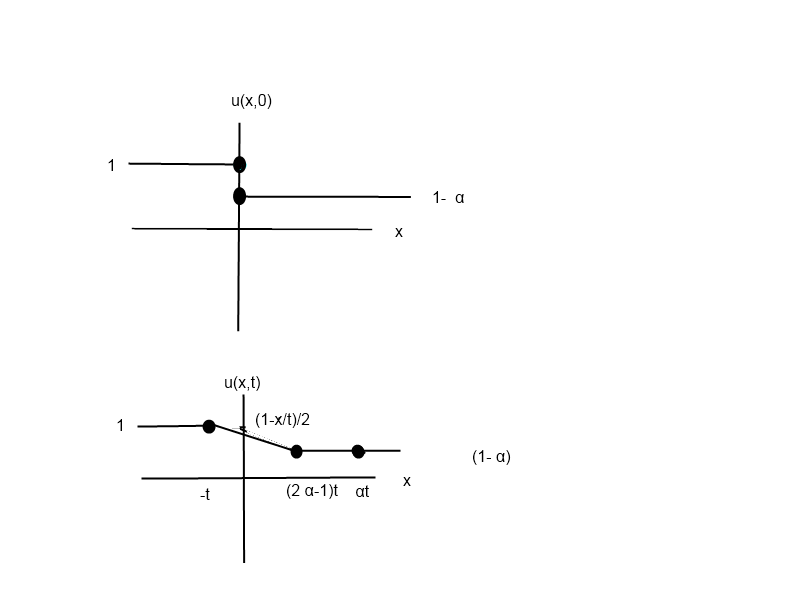}

The particle at position $\alpha t$ is the right most particle with rate $\alpha$.

\
By Burke's Theorem\cite{BP}, when there are $M$ particles with rate $\alpha$ placed on $\mathbb{N}$ initially, our initial condition is identical to step initial on $\mathbb{Z_{-}}$ and Bernoulli with density $(1-\alpha)$ on $\mathbb{Z_+}.$
Mathematically, the continuous version of $u(x,t)$, let's say $\rho(\xi,\tau)$ with $\rho(x/t,1)=u(x,t)$, satisfies the Burger's equation \begin{equation}\partial_\tau \rho+\partial_\xi(\rho(1-\rho))=0.\end{equation}

Since $u(x,t)=1-\alpha$ when $x>(2\alpha-1)t$ is a constant and there are $(1-\alpha)^2t$ particles moving around speed $\alpha$ , we know if $n\in(M,(1-\alpha)^2t)$, $n=$ area under $u(x,t)$ from $\overline{x}_n(t)$ to $\overline{x}_M(t)=\alpha t$ which is $n-M=\int_{\overline{x}_n}^{\overline{x}_M} u(x,t)dx$\footnote{Here we use $\mathbb{E}(x_n(t))=\overline{x}_n(t)$ as the notation for the average position of the particle $x_n(t)$ at time $t$.}

$$ n-M=(\alpha t-\overline{x}_n)(1-\alpha)$$
\begin{equation}\Rightarrow \overline{x}_{n}=\alpha t-(n-M)/(1-\alpha).\end{equation}







%

If we apply the above steps to the other regions and let $n/t=\nu$, then  we have
\begin{equation}
\lim_{t\to\infty}\frac{E(x_{[\nu t]}(t))}{t}:=
\left\{\begin{array}{ll}
\alpha-\nu/(1-\alpha),& \nu\in(0,(1-\alpha)^2),\\
1-2\sqrt{\nu},& \nu\in((1-\alpha)^2),1),\\
-\nu,& \nu\in(1,\infty).\\
\end{array}\right.
\end{equation}

Please be noted that we will always assume $0<\alpha<1.$

Since the particles get slowed down when $\nu\in[0,(1-\alpha)^2]$, we will look at fluctuations on a $\sqrt{t}$ scale in this region . For $\nu\in((1-\alpha)^2,(1))$, we will look at fluctuations on a $t^{1/3}$ scale, for $\nu > 1$, there is no fluctuation.

We now have the rescaled process given as
\begin{equation}
X_t (\nu):=
\left\{\begin{array}{ll}
\frac{x_{[\nu t]}-\{\alpha t-(\nu t-M)/(1-\alpha)\}}{-\sigma(\nu)\sqrt{t}},& \nu\in(0,(1-\alpha)^2), \sigma^2(\nu)=\alpha(1-\nu/(1-\alpha)^2)\\
\frac{x_{[\nu t]}-(t-2\sqrt{\nu t^2- Mt })}{-t^{1/3}},& \nu\in((1-\alpha)^2),1),\\
0,& \nu\in(1,\infty).\\
\end{array}\right.
\end{equation}

We now let $\pi(\theta)$ be a function from $\mathbb{R}\to \mathbb{R}$ such that $|\pi'|\leq 1.$
Define  $t_i=(\pi(\theta_i)+\theta_i)T$ , $n_i=(\pi(\theta_i)-\theta_i)T+M$ , $a=\pi-\theta$, and  $u=\pi+\theta$ then

\begin{equation}
X_t (\nu)=X_T(\theta)=
\left\{\begin{array}{ll}
\frac{x_{n(\theta,T)}(t(\theta,T))-(\alpha t-\frac{n-M}{1-\alpha})}{-\sigma\sqrt{T}},& \nu\in(0,(1-\alpha)^2), \sigma^2(\theta)=\alpha u-a/(1-\alpha)^2 \\
\frac{x_{[\nu t]}-(t-2\sqrt{\nu t^2- Mt })}{-T^{1/3}},& \nu\in((1-\alpha)^2),1),\\
0,& \nu\in(1,\infty).\\
\end{array}\right.
\end{equation}

Below is the process diagram which we will prove in this chapter .

\newpage
\textbf{Process Diagram:}\

\includegraphics[width=6in,height=3in]{p5.png}


In section 4, we will sketch the proof of convergence of the rescaled process $X_T$ in each region in the sense of finite dimensional distributions.


\section{Kernels}
\label{sec:LabelForKernels}

\subsection{Overview}
\label{sec:LabelForKernels:overview}
In this section, we briefly discuss about the correlation kernel for the joint probability distribution of the position for arbitrary number of particles.

Here we will write the joint probability distribution of positions $x_i(t)$ as a Fredholm determinant expression. Before doing that, we first consider a set of non-decreasing positive numbers $\{v_1,...,v_n\}$ such that $v_i\leq v_j$ if $i\leq j.$ Let $\{u_1<u_2<...<u_\nu\}$ be their different values with $\alpha_k$ being the multiplicity of $u_k.$ Also we define a set called space like set $$S=\{(n_k,t_k),k=1,...,m| n_k\geq n_{k+1}, t_k\leq t_{k+1}\}.$$  Define
\begin{equation}V_n=\{x^lu_k^x| 1\leq k\leq\nu,0\leq l\leq \alpha_k-1\}\end{equation} as a space of functions in $x$.  \\

\textbf{Theorem 3.1.1$\cite{Bo0}$:}\

Consider particles such that $x_i(0)=y_i.$ Take a sequence of particles and times in the space like set $S$. The joint probability distribution of $x_{n_k}(t_k)$ is given by the following Fredholm determinant expression
 \footnote{Another way to see Fredholm determinant is $$\det(1-\chi_aK\chi_a)_{l^2((\{(n_1,t_1),...,(n_m,t_m)\}) \times \mathbb{Z})}$$$$=\sum_{n\geq 0} \frac{(-1)^n}{n!}\sum_{i_1,...,i_n=1}^m\sum_{x_1\geq a_1}\cdots\sum_{x_n\geq a_n} \det K((n_{i_k},t_{i_k},x_k);(n_{i_j},t_{i_j},x_j))|_{1\leq j,k\leq n}$$ $$=\sum_{n\geq 0} \frac{(-1)^n}{n!}\sum_{i_1,...,i_n=1}^m\int_{\{y_1\geq a_1\}}dy_1\cdots\int_{\{y_n\geq a_n\}}dy_n \det K((n_{i_k},t_{i_k},[y_k]);(n_{i_j},t_{i_j},[y_j]))|_{1\leq j,k\leq n}$$ if $[y_i]=\text{Lowest integer greater or equal to } y_i.$}

\begin{equation}P(\cap_{k=1}^m\{x_{n_k}(t_k)\geq a_k\})=\det(1-\chi_aK\chi_a)_{\{l^2(\{(n_1,t_1),...,(n_m,t_m)\}\times \mathbb{Z} )\}}\end{equation}

where $\chi_a((n_k,t_k),x)=\mathbb{I}_{\{x<a_k\}}.$ Here $K$ is a kernel on $l^2 (S \times \mathbb{Z})$ and it is defined as the following

\begin{equation}
K((n_1,t_1),x_1;(n_2,t_2),x_2)  =-\phi^{((n_1,t_1),(n_2,t_2))}(x_1,x_2)+\sum_{k=1}^{n_2}\Psi^{n_1,t_1}_{n_1-k}(x_1)\Phi^{n_2,t_2}_{n_2-k}(x_2)
\end{equation}
where
  \begin{equation}\Psi^{n,t}_{n-j}(x)=\frac{1}{2\pi i}\oint_{\Gamma_{0,\vec{v}}} \frac{dw}{w} \frac{e^{tw}}{w^{x-y_j+n-j}}\frac{\prod_{k=1}^{n}(w-v_k)}{\prod_{k=1}^{j}(w-v_j)}.\end{equation}
and   \begin{equation}\phi^{((n_1,t_1),(n_2,t_2)}(x_1,x_2)=\frac{1}{2\pi i}\oint_{\Gamma_{0,\vec{v}}} \frac{dw}{w} \frac{e^{(t_1-t_2)w}}{w^{x+n_1-x_2-n_2}}\frac{\mathbb{I}_{S}}{(w-1)^{n_2-n_1}}\end{equation} where $\mathbb{I}_{S}$ is the identity function on the space like space S , $\vec{v}=\{v_{n_1+1},...,v_{n_2}\}$ and $\Gamma_{0,\vec{v}}$ denotes any positive oriented simple closed curve that includes $0$ and the points in the set $\vec{v}$.

The functions $\{\Phi^{n,t}_{n-j}\}_{1\leq j\leq n}$ are described by the following conditions:

  \begin{equation}
<\Phi^{n,t}_{n-j},\Psi^{n,t}_{n-k}>=\delta_{j,k}, \ 1\leq j,k,\leq n,
\end{equation}

and span$\{\Phi_{n-j}^{n,t}(x),1\leq j\leq n\}=V_n.$ \\

%
%
The orthornormal polynomials for step initial condition $y_j=M-j$ are given at next section.
\newpage
\subsection{Bi-orthornormal polynomials}
\label{sec:LabelForKernels:Bi-orthornormal polynomials}
In this section we verify the orthogonal condition for our kernel with step initial condition (see picture below).\

\includegraphics[width=5in,height=1in]{p2-1.png}\

Orange particles have hopping rate $\alpha$ and black particles have hopping rate $1.$

The main difficulty to apply Theorem 3.1.3 is to find the corresponding bi-orthogonal polynomials for its kernel. It was pointed in \cite{Bo2} that bi-orthogonally polynomials in \cite{Bo2} for the case of alternating initial condition can be obtained by Gram-Schmidt orthogonalization procedure. However, we do not do it here. We obtained our bi-orthogonal polynomials by modifying the bi-orthogonal polynomials for the case of alternating initial condition.
They are determined once the bi-orthogonal condition is meet.

We will show that the $\Phi^{n,t}_{n-j}(x)$ functions which we will define later satisfies the following relationship
 \begin{equation}
<\Phi^{n,t}_{n-j},\Psi^{n,t}_{n-k}>:=\sum_{x\in\mathbb{Z}}\Phi^{n,t}_{n-j}(x)\Psi^{n,t}_{n-k}(x)= \delta_{j,k}, \ 1\leq j,k,\leq n.
\end{equation}

$\Phi^{n,t}_{n-j}$ and $\Psi^{n,t}_{n-k}$ are called bi-orthogonal if they all satisfy the above relationship.
The above $<,>$ map can be viewed as an inner product function from $V_n\times V_n\to \mathbb{Z}.$ Using the Cauchy integral formula, one can see the $\Phi^{n,t}_{n-j}(x)$ functions span $V_n.$ Thus, together with the following theorem,  $\Phi^{n,t}_{n-j}$ will satisfy the condition for Theorem 3.1.1.

From now on, every contour integral's \footnote{Instead of writing $\oint \frac{dz}{2\pi i}$, we write $\oint dz.$} differential $dz$ is identified as $\frac{dz}{2\pi i}.$

In previous section, we know our kernel is expressed as a pair of bi-orthogonal polynomials $\Phi$ and $\Psi$, our polynomials is constructed based on the polynomials in $\cite{Bo2}$. We will follow their notations as much as possible. With step initial condition $y_i=M-i$ and letting the first $M$ particles have rate $\alpha$ and the rest of particles with unit speed(see figure above for an example with $M=3$), the orthonormal polynomials for the correlation kernel in previous section are given as the following which is slight different from the case of alternating initial condition $\cite{Bo2}$. \


\textbf{Theorem 3.2.1 Bi-orthonormal polynomials for the Kernel $K$:}\

From now on we use the notation $\Gamma_{p_1,...,p_l}$ to represent a positive oriented contour integral enclosed the points: $p_1,...,p_n.$\

(a) For $n\leq M$,
\begin{equation}\Psi^{n,t}_{n-j}(x)=\frac{1}{2\pi i}\oint_{\Gamma_{0,1}}\frac{dw}{w}\frac{(w-\alpha)^{n-j}}{w^{x+n-M}}e^{tw}\end{equation}

\begin{equation}\Phi^{n,t}_{n-j}(x)=\frac{1}{2\pi i}\oint_{\Gamma_{\alpha-1}}dv\frac{(1+v)^{x+n-M}}{(v-(\alpha-1))^{n-j+1}}e^{-t(v+1)}.\end{equation}

(b) For $n\geq M$ and $j\geq M+1$,

\begin{equation}\Psi^{n,t}_{n-j}(x)=\frac{1}{2\pi i}\oint_{\Gamma_{0,1}}\frac{dw}{w}\frac{(w-1)^{n-j}}{w^{x+n-M}}e^{tw}\end{equation}
\begin{equation}\Phi^{n,t}_{n-j}(x)=\frac{1}{2\pi i}\oint_{\Gamma_{0}}dv\frac{(1+v)^{x+n-M}}{v^{n-j+1}}e^{-t(v+1)}.\end{equation}

(c) For $n\geq M$ and $j\leq M$,

\begin{equation}\Psi^{n,t}_{n-j}(x)=\frac{1}{2\pi i}\oint_{\Gamma_{0,\alpha}}\frac{dw}{w}\frac{(w-\alpha)^{M-j}(w-1)^{n-M}}{w^{x+n-M}}e^{tw}\end{equation}
\begin{equation}\Phi^{n,t}_{n-j}(x)=\frac{1}{(2\pi i)^2}\oint_{\Gamma_{\alpha-1}}dv \frac{1}{(v-(\alpha-1))^{M-j+1}}\oint_{\Gamma_{0,v}}dz \frac{1}{z-v} \frac{(1+z)^{x+n-M}}{z^{n-M}e^{t(z+1)}}g(z,v).\end{equation}


Here,
\begin{equation}g(z,v)=\frac{2z+A}{z+v+A}\frac{(z+A)^{M-j}}{(v+A)^{M-j}},\end{equation} and $A>>|z|, z\in\Gamma_{0,v}.$

\textbf{Proof:}\


First we recall by definition $$<\Psi^{n,t}_{n-j},\Phi^{n,t}_{n-k}>=\sum_{x\in \mathbb{Z}}\Psi^{n,t}_{n-j}(x)\Phi^{n,t}_{n-k}(x)$$ and since $\Psi^{n,t}_{n-j}(x)=0$ if $x\leq M-n-1$ we have \begin{equation}\sum_{x\in \mathbb{Z}}\Psi^{n,t}_{n-j}(x),\Phi^{n,t}_{n-k}(x)=\sum_{x\geq M-N }\Psi^{n,t}_{n-j}(x)\Phi^{n,t}_{n-k}(x).\end{equation}
The proof is by direct computation and the fact that $$\sum_{x\geq M-n}(\frac{1+v}{w})^{x+n-M}=\frac{w}{w-(1+v)}.$$  There are four cases to prove for $n\geq M+1$ and one case for $n\leq M$.

When $n\leq M$, direct computation shows the following if we require $|\frac{1+v}{w}|<1$ then $\sum_{x\geq M-n}\frac{(1+v)^{x+n-M}}{w^{x+n-M}}=\frac{w}{w-(1+v)}$
$$<\Psi^{n,t}_{n-j},\Phi^{n,t}_{n-k}>=\frac{1}{(2\pi i)^2}\oint_{\Gamma_{0,1}}dw\oint_{\Gamma_{\alpha-1}}dv \frac{dw}{w}\frac{(w-\alpha)^{n-j}}{(v+1-\alpha)^{n-k+1}}\frac{e^{tw}}{e^{t(v+1)}}\frac{w}{w-(1+v)}$$

and if we evaluate the only pole at $w=1+v$\footnote{The pole at $w=0$ does not exist anymore after we summing up x. } we have

$$=\frac{1}{2\pi i} \oint_{\Gamma_{\alpha-1}} \frac{1}{(v+1-\alpha)^{j-k+1}} $$

then let $z=v+1-\alpha$

$$=\frac{1}{2\pi i} \oint_{\Gamma_{0}} \frac{1}{z^{j-k+1}}=\delta_{jk}.$$


When $n\geq M+1$,

The cases for $(i)$ $j\leq M$, $k\leq M$ , $(ii)$ $j\geq M+1$, $k\geq M+1$, $(iii)$ $j\geq M+1$, $k\geq M$ are all similar to the first calculation.


$(i)$ $j\leq M$, $k\leq M:$

Since we have $\Psi^{n,t}_{n-j}(x)=0$ if $x\leq M-n-1$,
$$<\Psi^{n,t}_{n-j},\Phi^{n,t}_{n-k}>=\sum_{x\in \mathbb{Z} }\Psi^{n,t}_{n-j}(x)\Phi^{n,t}_{n-k}(x)=\sum_{x\geq M-N }\Psi^{n,t}_{n-j}(x)\Phi^{n,t}_{n-k}(x)$$

and after summing up $\sum_{x\geq M-n}(\frac{1+z}{w})^{x+n-M}=\frac{w}{w-(1+z)}$  by requiring $|\frac{1+z}{w}|<1$ and evaluating the only pole at $w=z+1$ then
$$=\frac{1}{(2\pi i)^2}\oint_{\Gamma_{\alpha-1}}dv\oint_{\Gamma_{0,v}}dz \frac{(z+1-\alpha)^{M-j}}{z-v}\frac{1}{(v+1-\alpha)^{M-k+1}}g(z,v)$$
and now we evaluate the pole at $z=v$
$$=\frac{1}{2\pi i}\oint_{\Gamma_{\alpha-1}} \frac{1}{(v+1-\alpha)^{j-k+1}}=\delta_{j,k}.$$


$(ii)$ $j\geq M+1$, $k\geq M+1:$

Similarly to previous proofs we have

$$<\Psi^{n,t}_{n-j},\Phi^{n,t}_{n-k}>=\sum_{x\in \mathbb{Z} }\Psi^{n,t}_{n-j}(x)\Phi^{n,t}_{n-k}(x)=\sum_{x\geq M-N }\Psi^{n,t}_{n-j}(x)\Phi^{n,t}_{n-k}(x)$$ since $\Psi^{n,t}_{n-j}(x)=0$ if $x\leq M-n-1$,

and after summing up $\sum_{x\geq M-n}(\frac{1+v}{w})^{x+n-M}=\frac{w}{w-(1+v)}$  by requiring $|\frac{1+v}{w}|<1$ and evaluating the only pole at $w=v+1$ then

$$=\frac{1}{2\pi i}\oint_{\Gamma_{0}}dv  \frac{1}{v^{j-k+1}}=\delta_{j,k}.$$

$(iii)$ $j\leq M$, $k\geq M+1:$
Since we have $\Psi^{n,t}_{n-j}(x)=0$ if $x\leq M-n-1$,
$$<\Psi^{n,t}_{n-j},\Phi^{n,t}_{n-k}>=\sum_{x\in \mathbb{Z} }\Psi^{n,t}_{n-j}(x)\Phi^{n,t}_{n-k}(x)=\sum_{x\geq M-N }\Psi^{n,t}_{n-j}(x)\Phi^{n,t}_{n-k}(x)$$

and after summing up $\sum_{x\geq M-n}(\frac{1+v}{w})^{x+n-M}=\frac{w}{w-(1+v)}$  by requiring $|\frac{1+v}{w}|<1$ and evaluating the only pole at $w=v+1$ then

$$=\frac{1}{2\pi i}\oint_{\Gamma_{0}}\frac{1}{v^{M-k+1}}(v+1-\alpha)^{M-j}=0.$$


$(iv)$ $j \geq M+1$ , $k\leq M$ is different.

Now we require $$|\frac{1+z}{w}|<1$$ then after summing up $\sum_{x\geq M-n}(\frac{1+z}{w})^{x+n-M}=\frac{w}{w-(1+z)}$ and substituting the pole $w=z+1$ we have

$$<\Psi^{n,t}_{n-j},\Phi^{n,t}_{n-k}>=\frac{1}{(2\pi i)^2}\oint_{\Gamma_{\alpha-1}}dv \oint_{\Gamma_{0,v}}dz \frac{1}{z-v}\frac{z^{n-j}}{z^{n-M}}g(z,v),$$

then it's not hard to see that\footnote{Recall $$g(z,v)=\frac{2z+A}{z+v+A}\frac{(z+A)^{M-j}}{(v+A)^{M-j}}$$ has poles at $-A$ and $-A-v,$ where $A$ is a positive constant such that $-A-v$ is not a pole inside the contour of $\Gamma_{0,v}$ in the $z$ plane. } $$dz \frac{1}{z-v}\frac{z^{n-j}}{z^{n-M}}g(z,v) =dw \frac{1}{w-v}\frac{w^{n-j}}{w^{n-M}}g(w,v)$$ if we do the substitution $z=-w-A.$

Therefore $$\frac{1}{2\pi i}\oint_{\Gamma_{0,v}}dz \frac{1}{z-v}\frac{z^{n-j}}{z^{n-M}}g(z,v)=\frac{1}{2\pi i}\oint_{\Gamma_{-A,-A-v}}dw \frac{1}{w-v}\frac{w^{n-j}}{w^{n-M}}g(w,v)$$
$$=\frac{1}{4\pi i}\oint_{\Gamma_{0,v,-A,-A-v}}dz \frac{1}{z-v}\frac{z^{n-j}}{z^{n-M}}g(z,v).$$ Now the integrand is of order at most $O(\frac{1}{z^{3}})$ as $z\to\infty$ since $j-M\geq 1$ and therefore the integral in $z$ plane is $0$. So, $$<\Psi^{n,t}_{n-j},\Phi^{n,t}_{n-k}>=0.$$$\square$
\newpage
\subsection{Construction of Kernel}
\label{sec:LabelForKernels:Construction of Kernel}
Now we are ready to write out our kernel and again for simplicity, we use $\Gamma_{p_1,...,p_l}$ to represent a positive oriented contour enclosed the points: $p_1,...,p_l$.


From previous section, we know $$K((n_1,t_1),x_1;(n_2,t_2),x_2)  =-\phi^{((n_1,t_1),(n_2,t_2))}(x_1,x_2)+\sum_{k=1}^{n_2}\Psi^{n_1,t_1}_{n_1-k}(x_1)\Phi^{n_2,t_2}_{n_2-k}(x_2).$$
For $n_1$ and $n_2\geq M+1$, the kernel can be rewritten as

$$K((n_1,t_1),x_1;(n_2,t_2),x_2)=-\phi^{((n_1,t_1),(n_2,t_2))}(x_1,x_2)+\sum_{k=1}^{M}\Psi^{n_1,t_1}_{n_1-k}(x_1)\Phi^{n_2,t_2}_{n_2-k}(x_2)$$$$+\sum_{k=M+1}^{n_2}\Psi^{n_1,t_1}_{n_1-k}(x_1)\Phi^{n_2,t_2}_{n_2-k}(x_2)$$
$$=-\phi^{(n_1,t_1),(n_2,t_2)}(x_1,x_2)+ K_1((n_1,t_1),x_1;(n_2,t_2),x_2)+K_2((n_1,t_1),x_1;(n_2,t_2),x_2).$$
\begin{equation}\end{equation}
where $$\phi^{((n_1,t_1),(n_2,t_2))}(x_1,x_2)=\frac{1}{2\pi i}\oint_{\Gamma_{0,1}} \frac{dw}{w} \frac{e^{(t_1-t_2)w}}{w^{x_1+n_1-x_2-n_2}}\frac{\mathbb{I}_{S}}{(w-1)^{n_2-n_1}}.$$, $$K_1((n_1,t_1),x_1;(n_2,t_2),x_2)=\sum_{j=M+1}^{n_2} \Psi^{n,t}_{n-j}(x_1)\Phi^{n,t}_{n-j}(x_2)$$ and $$K_2((n_1,t_1),x_1;(n_2,t_w),x_2)=\sum_{j=1}^{M} \Psi^{n_1,t}_{n_1-j}(x_1)\Phi^{n_2,t}_{n_2-j}(x_2).$$

\textbf{Theorem 3.3.1: }\

We can also simplify the expression of $\phi$, $K_1$ and $K_2$ which we will denote the simplified version of them as $\hat{\phi}$,$\hat{K_1}$ and $\hat{K_2},$ respectively. As a result we have the following

\begin{equation}K((n_1,t_1),x_1;(n_2,t_2),x_2)=-\hat{\phi}^{(n_1,t_1),(n_2,t_2)}(x_1,x_2)+\end{equation} $$\hat{K_1}((n_1,t_1),x_1;(n_2,t_2),x_2)+\hat{K_2}((n_1,t_1),x_1;(n_2,t_2),x_2)$$ where
\begin{equation}\hat{\phi}^{(n_1,t_1),(n_2,t_2)}(x_1,x_2)=\frac{1}{2\pi i}\oint_{\Gamma_{0}} \frac{dw}{w} \frac{e^{(t_1-t_2)w}}{w^{x_1+n_1-x_2-n_2}}\frac{\mathbb{I}_{S}}{(w-1)^{n_2-n_1}},\end{equation}
\begin{equation}\hat{K_1}((n_1,t_1),x_1;(n_2,t_2),x_2)=\frac{1}{(2\pi i)^2}\oint_{\Gamma_1} dv\oint_{\Gamma_{0}} dw\frac{(w-1)^{n_1-M}}{w^{x+n_1-M+1}}\frac{e^{t_1w}}{e^{t_2v}}\frac{v^{x_2+n_2-M}}{(v-1)^{n_2-M}}\frac{1}{w-v},\text{and}\end{equation}

$\hat{K}_2((n_1,t_1),x_1;(n_2,t_w),x_2)=$ $$\frac{1}{(2\pi i)^3}\oint_{\Gamma_{\alpha}}dv\oint_{\Gamma_{1,v}}dz   \oint_{\Gamma_{0}}\frac{dw}{w} \frac{e^{t_1w} (w-1)^{n_1-M}}{w^{x_1+n_1-M}} \frac{z^{x_2+n_2-M}}{(z-1)^{n_2-M}e^{t_2z}}\frac{(w-\alpha)^M}{(v-\alpha)^M}\frac{2(z-1)+A}{z-1+v-1+A}\frac{1}{z-v}\cdot$$
 $$\frac{(z-1+A)^M}{(v-1+A)^M}\frac{v-1+A}{(z-1+A)(w-\alpha)-(v-1+A)(v-\alpha)}. $$
\begin{equation}\end{equation}
Here we require $|v|<|w-1|$ for $\hat{K}_1$ and $|v-\alpha+1|<|w-\alpha|$ for $\hat{K}_2.$\

\textbf{Proof:}\

The proof is by summing up geometric series.
$$\hat{K}_1((n_1,t_1),x_1;(n_2,t_2),x_2)=\sum_{j=M+1}^{n_2} \Psi^{n,t}_{n-j}(x_1)\Phi^{n,t}_{n-j}(x_2).$$

Since $\phi^{((n_1,t_1),(n_2,t_2))}(x_1,x_2)=0$ if $k\geq n_2+1$, we can extend the sum from $n_2$ to $\infty.$
So we have

$$\hat{K}_1=\frac{1}{(2\pi i)^2}\oint_{\Gamma_0}dv \oint_{\Gamma_{0,1}}\frac{dw}{w}\frac{e^{t_1w}(w-1)^{n_1}}{w^{x_1+n_1-M}}\frac{(1+v)^{x_2+n_2-M}}{v^{n_2+1}e^{t_2(v+1)}}\sum_{k=M+1}^\infty\frac{v^k}{(w-1)^k}$$

$$=\frac{1}{(2\pi i)^2}\oint_{\Gamma_0}dv \oint_{\Gamma_{0,1}}\frac{dw}{w}\frac{e^{t_1w}(w-1)^{n_1}}{w^{x_1+n_1-M}}\frac{(1+v)^{x_2+n_2-M}}{v^{n_2+1}e^{t_2(v+1)}}\frac{v^{M+1}}{(w-1)^{M}(w-1-v)}.$$  \begin{equation}\end{equation} Here we require $|v|<|w-1|$ which can be achieved if we make $v$ small enough.

After change of variable $v\to v-1$ we have our

$$\hat{K}_1=\frac{1}{(2\pi i)^2}\oint_{\Gamma_1}dv \oint_{\Gamma_{0,1}}\frac{dw}{w}\frac{e^{t_1w}(w-1)^{n_1}}{w^{x_1+n_1-M}}\frac{v^{x_2+n_2-M}}{(v-1)^{n_2+1}e^{t_2 v}}\frac{(v-1)^{M+1}}{(w-1)^{M}(w-v)}.$$

Now we proceed to $\hat{K}_2.$

$$\hat{K}_2((n_1,t_1),x_1;(n_2,t_w),x_2)=\sum_{j=1}^{M} \Psi^{n_1,t}_{n_1-j}(x_1)\Phi^{n_2,t_2}_{n-j}(x_2)$$
$$=\frac{1}{2\pi i}\oint_{\Gamma_{0,\alpha}}\frac{dw}{w}\frac{e^{t_1w} (w-1)^{n_1-M}}{w^{x_1+n_1-M}}\times$$ $$\frac{1}{(2\pi i)^2}\oint_{\Gamma_{\alpha-1}}\frac{dv}{v-\alpha+1}   \oint_{\Gamma_{0,v}}dz \frac{(1+z)^{x_2+n_2-M}}{z^{n_2-M}e^{t_2(z+1)}}\frac{(w-\alpha)^M}{(v+1-\alpha)^M}\frac{2z+A}{z+v+A}\frac{1}{z-v}\cdot$$
 $$\frac{(z+A)^M}{(v+A)^M}\sum_{k=1}^\infty\frac{(v+A)^{k}(v+1-\alpha)^k}{(w-\alpha)^k(z+A)^k}. $$
After summing up the series and make $z$ , $w$ big enough and $v$ close to $\alpha-1$ enough such that \begin{equation}|v+1-\alpha||v+A|<|z+A||w-\alpha|.\end{equation} we have the following

$$\hat{K}_2((n_1,t_1),x_1;(n_2,t_w),x_2)=\frac{1}{(2\pi i)^3}\oint_{\Gamma_{0,\alpha}}\frac{dw}{w}\frac{e^{t_1w} (w-1)^{n_1-M}}{w^{x_1+n_1-M}}\times$$ $$\oint_{\Gamma_{\alpha-1}}\frac{dv}{v-\alpha+1}   \oint_{\Gamma_{0,v}}dz \frac{(1+z)^{x_2+n_2-M}}{z^{n_2-M}e^{t_2(z+1)}}\frac{(w-\alpha)^M}{(v+1-\alpha)^M}\frac{2z+A}{z+v+A}\frac{1}{z-v}\cdot$$
 $$\frac{(z+A)^M}{(v+A)^M}\frac{(v+A)(v+1-\alpha)}{(z+A)(w-\alpha)-(v+A)(v+1-\alpha)}. $$

\begin{equation}\end{equation}

If we compute the simple pole at $w=\alpha+\frac{(v-\alpha)(v+A-1)}{(z-1+A)}$ we have $(z+A)(w-\alpha)=(v+A)(v+1-\alpha),$ then the residue is zero since the contour integral on the $v$ plane is zero.

Now for convenience which we will see later, we change the order of integration along with the change of variables $v\to v-1$ and $z\to z-1$ then we have

$$\hat{K}_2((n_1,t_1),x_1;(n_2,t_w),x_2)=\frac{1}{(2\pi i)^3}\oint_{\Gamma_{\alpha}}\frac{dv}{v-\alpha}\times$$ $$\oint_{\Gamma_{1,v}}dz   \oint_{\Gamma_{0}}\frac{dw}{w} \frac{e^{t_1w} (w-1)^{n_1-M}}{w^{x_1+n_1-M}} \frac{z^{x_2+n_2-M}}{(z-1)^{n_2-M}e^{t_2z}}\frac{(w-\alpha)^M}{(v-\alpha)^M}\frac{2(z-1)+A}{z-1+v-1+A}\frac{1}{z-v}\cdot$$
 $$\frac{(z-1+A)^M}{(v-1+A)^M}\frac{(v-1+A)(v-\alpha)}{(z-1+A)(w-\alpha)-(v-\alpha)(v-1+A)}. $$
\begin{equation}\end{equation}
$\square$

%

\section{Asymptotics}
\label{sec:LabelForAsymptotics:}
In the next three sections, we will sketch the proof of convergence of the rescaled process $X_T$ in each region in the sense of finite dimensional distributions. We will follow the procedures described in \cite{Bo4}.

\subsection{Slow Particles Region}

\textbf{Theorem 4.1.1}\

Here we will show

\begin{equation} \lim_{t\to\infty}X_t(\nu)=\text{DBM}(-\ln(\sigma(v))), \nu\in(0,(1-\alpha)^2)\end{equation}
in the sense of finite dimensional distributions.
It should be noted here our results is similar to \cite{Bo2}.

To prove it, we prove a more general version of Theorem 3.4.1.

\textbf{Proposition 4.1.2.}\

We let $\pi(\theta)$ be a function from $\mathbb{R}\to \mathbb{R}$ such that $|\pi'|\leq 1.$
Define  $t_i=(\pi(\theta_i)+\theta_i)T$ and $n_i=(\pi(\theta_i)-\theta_i)T+M$ then for $0<n<(1-\alpha^2)t$ which means $0<\pi(\theta)<\frac{\alpha^2-2\alpha+2}{\alpha(2-\alpha)}\theta,$ we have

\begin{equation}X_T(\theta)=\frac{x_{n(\theta,T)}(t(\theta,T))-(\alpha t-\frac{n-M}{1-\alpha})}{-\sigma\sqrt{T}} \to \text{DBM}(\tau(\theta))\ \text{as}\ T\to\infty\end{equation}

with $\sigma^2:=\alpha(\pi+\theta)-\frac{\alpha(\pi-\theta)}{(1-\alpha)^2}$ and $\tau(\theta)=-\ln(\sigma).$

\textbf{Proof}:\

We need to show the kernel $$K((n_1,t_1),x_1;(n_2,t_2),x_2)=-\hat{\phi}^{(n_1,t_1),(n_2,t_2)}(x_1,x_2)+[\hat{K_1}+\hat{K_2}]((n_1,t_1),x_1;(n_2,t_2),x_2)$$ converges to $K^{{\rm DBM}}$ after rescaling. Here we have


\begin{equation}x_i=\alpha t_i-n_i/(1-\alpha)-\xi_i\sigma_i\sqrt{T}, \end{equation}
\begin{equation}u_i=\pi(\theta_i)+\theta_i, \end{equation}
\begin{equation}a_i=\pi(\theta_i)-\theta_i, \end{equation}
\begin{equation}\sigma_i^2=\alpha u_i-\alpha a_i/(1-\alpha)^2. \end{equation}
\begin{equation}t_i=u_iT,\end{equation}
\begin{equation}n_i=M+a_iT.\end{equation}

Since we require $t_1\geq t_2$ if $n_2\geq n_1$, we have $a_2-a_1\geq 0$ and $u_1-u_2\geq 0$ with one of them being strict inequality.

There are three parts to show.


\begin{equation}
\begin{aligned}
&\lim_{T\to\infty} C \hat{\phi}= \frac{\exp\left(-{\displaystyle \frac{(\xi_1-\xi_2\sigma_2/\sigma_1)^2}{2(1-\sigma_2^2/\sigma_1^2))}}\right)}{\sqrt{2\pi(1-\sigma_2^2/\sigma_1^2)}},\\
&\lim_{T\to\infty} C K_1=0,\\
&\lim_{T\to\infty} C K_2= \oint_{|V|=\mathbb{R}}dV\oint_{-L+i\mathbb{R}}dW\frac{W^M}{V^M}\frac{1}{W-V}\frac{e^{W^2/2+W\xi_1}}{e^{V^2(\sigma_2/\sigma_1)^2+V\xi_2\sigma_2/\sigma_1}}.
\end{aligned}
\end{equation}

with $L>R$ and \begin{equation} C=\frac{e^{\alpha(t_2-t_1)}(\alpha-1)^{n_2-n_1}}{\alpha^{x_2-x_1+n_2-n_1}}\sigma_1\sqrt{T}. \end{equation}

\textbf{Analysis of }$\hat{\phi}$:\

First, we define the following functions

\begin{equation}
\begin{aligned}
g_0(w)&=(u_1-u_2)(w-\alpha\ln(w))+(a_1-a_2)(\ln(w-1)+\frac{\alpha}{1-\alpha}\ln(w)),\\
g_1(w)&=(\xi_1\sigma_1-\xi_2\sigma_2)\ln(w),\\
g_2(w)&=-\ln(w).
\end{aligned}
\end{equation}

In order to do steepest descent, we must calculate their derivatives and we found

\begin{equation}
\begin{aligned}
g_0'(\alpha)&=0,\\
g_0''(\alpha)&=\frac{u_1-u_2}{\alpha}+\frac{a_2-a_1}{(1-\alpha)^2\alpha}>0,\\
g_1'(\alpha)&=\frac{\xi_1\sigma_1-\xi_2\sigma_2}{\alpha},\\
g_2'(\alpha)&=-\frac{1}{\alpha}.
\end{aligned}
\end{equation}

Using $w-\alpha=\alpha e^{iy}-\alpha=\alpha(iy+\frac{1}{2}(iy)^2+...)$ their Taylor expansion at $w=\alpha$ are given as

\begin{equation}
\begin{aligned}
g_0(w)&=g_0(\alpha)-\frac{y^2}{2}(\sigma_1^2-\sigma_2^2)+O(y^3),\\
g_1(w)&=g_1(\alpha)+iy(\xi_1\sigma_1-\xi_2\sigma_2)+O(y^2),\\
g_2(w)&=-\ln(\alpha)+O(y).
\end{aligned}
\end{equation}

Let $\Gamma_0=\{\alpha e^{iy}, y\in(-\pi,\pi]\}$ and use $e^x\sim 1+ x+O(x^2)$ for $x<<1$ then for some $\delta>0$ we have

\begin{equation} \begin{aligned}
\hat{\phi}&=\oint_{\Gamma_0} dw e^{Tg_0(w)+\sqrt{T}g_1(w)+g_2(w)}\\
	  &=\oint_{ |w-\alpha|<\delta} dw e^{Tg_0(w)+\sqrt{T}g_1(w)+g_2(w)}+\oint_{\Gamma_0 \backslash \{|w-\alpha|<\delta\}} dw e^{Tg_0(w)+\sqrt{T}g_1(w)+g_2(w)}\\
	   &=\oint_{ |w-\alpha|<\delta} dw e^{Tg_0(w)+\sqrt{T}g_1(w)+g_2(w)}+e^{Tg_0(\alpha)+\sqrt{T}g_1(\alpha)+g_2(\alpha)}
\oint_{\Gamma_0 \backslash \{|w-\alpha|<\delta\}} dw e^{-\mu  T},
\end{aligned}
\end{equation}
After multipling by $C$ the second term has an error of order $O(e^{-\mu T})$ and $\mu\sim \delta^2.$ After we drop the error term, we have
\begin{equation} \begin{aligned}
          &=\oint_{\Gamma_0\cap \{|w-\alpha|<\delta\}} dw e^{T(g_0(\alpha)-\frac{1}{2}y^2(\sigma_1^2-\sigma^2_2) )+\sqrt{T}(g_1(\alpha)+iy(\xi_1\sigma_1-\xi_2\sigma_2))-\ln(\alpha)+TO(y^3)+\sqrt{T}O(y^2)+O(y)}\\
          &=  e^{Tg_0(\alpha) +\sqrt{T}g_1(\alpha) -\ln(\alpha)}\oint_{\Gamma_0\cap \{|w-\alpha|<\delta\}} dw e^{-T\frac{1}{2}y^2(\sigma_1^2-\sigma^2_2) +iy\sqrt{T}(\xi_1\sigma_1-\xi_2\sigma_2) }\times e^{TO(y^3)+\sqrt{T}O(y^2)+O(y)}\\
	  &=  e^{Tg_0(\alpha) +\sqrt{T}g_1(\alpha) -\ln(\alpha)}\oint_{\Gamma_0\cap \{|w-\alpha|<\delta\}} dw e^{-T\frac{1}{2}y^2(\sigma_1^2-\sigma^2_2) +iy\sqrt{T}(\xi_1\sigma_1-\xi_2\sigma_2) }\\
&+ e^{Tg_0(\alpha) +\sqrt{T}g_1(\alpha) -\ln(\alpha)}\oint_{\Gamma_0\cap  \{|w-\alpha|<\delta\}} dw e^{-T\frac{1}{2}y^2(\sigma_1^2-\sigma^2_2) +iy\sqrt{T}(\xi_1\sigma_1-\xi_2\sigma_2) }(e^{TO(y^3)+\sqrt{T}O(y^2)+O(y)}-1 ).
\end{aligned}
\end{equation}

Now we use $|e^x-1|\leq e^{|x|}|x|,$ then we have the second term of the above equation becomes
\begin{equation} \begin{aligned}
&\text{cnst}\cdot |\oint_{\Gamma_0\cap \{|w-\alpha|<\delta\}}dy e^{-T\frac{1}{2}y^2(\sigma_1^2-\sigma^2_2) +iy\sqrt{T}(\xi_1\sigma_1-\xi_2\sigma_2) }(e^{TO(y^3)+\sqrt{T}O(y^2)+O(y)}-1 )|\\
&\leq \text{cnst}\int_{0}^\delta dy |e^{-T\frac{1}{2}y^2(\sigma_1^2-\sigma^2_2) +iy\sqrt{T}(\xi_1\sigma_1-\xi_2\sigma_2) }(e^{TO(y^3)+\sqrt{T}O(y^2)+O(y)}-1 )|\\
&\leq \text{cnst}\int_{0}^\delta dy |e^{-T\frac{1}{2}y^2(\sigma_1^2-\sigma^2_2) +iy\sqrt{T}(\xi_1\sigma_1-\xi_2\sigma_2) }||e^{O(Ty^3+\sqrt{T}y^2+y)}O(y^3+\sqrt{T}y^2+y))|.
\end{aligned}
\end{equation}

We do the change of variable $p=y \sqrt T$, then the above becomes

$$=\frac{\text{cnst}}{T}\int_{0}^{\sqrt{T}\delta}  e^{-cp^2+... }O(p^3+...)dp.$$

After mutipling by $\sqrt{T}$, the estimate of error is $O(1/\sqrt{T})$ which goes to $0$ as $T$ goes to $\infty.$

So now the last step is to calculate the first term in equation (4.1.22).
$$e^{Tg_0(\alpha) +\sqrt{T}g_1(\alpha) -\ln(\alpha)}\oint_{\Gamma_0\cap \{|w-\alpha|<\delta\}} dw e^{-T\frac{1}{2}y^2(\sigma_1^2-\sigma^2_2) +iy\sqrt{T}(\xi_1\sigma_1-\xi_2\sigma_2) }.$$

If we let $1/D=\frac{e^{t_1\alpha}(\alpha-1)^{n_1}}{e^{t_2\alpha}(\alpha-1)^{n_2}}\frac{\alpha^{x_2+n_2}}{\alpha^{x_1+n_1}}=e^{Tg_0(\alpha)+\sqrt{T}g_1(\alpha)}$ , $w=\alpha e^{iy}$, , $z/\sqrt{T}=y$ , and extend $\delta\to\infty$ for $w\in\Gamma_0^\delta=\{w\in\Gamma_0,|w-\alpha|<\delta, w-\alpha=re^{\pm i\pi/2}\}$ for some $r$ then

$$  \frac{1}{D} \oint_{\Gamma_0\cap \{|w-\alpha|<\delta\}} dw e^{-T\frac{1}{2}y^2(\sigma_1^2-\sigma^2_2) +iy\sqrt{T}(\xi_1\sigma_1-\xi_2\sigma_2) }$$
\begin{equation} \begin{aligned}
&=  1/(D\sqrt{T}) \int_{\mathbb{R}} dw e^{-\frac{1}{2}z^2(\sigma_1^2-\sigma^2_2) +iz(\xi_1\sigma_1-\xi_2\sigma_2) }+ 1/(D\sqrt{T})O(e^{-\delta^2 T}) \\
&=1/C \int_{\mathbb{R}} dw e^{-\frac{1}{2}z^2(\sigma_1^2-\sigma^2_2) +iz(\xi_1\sigma_1-\xi_2\sigma_2) }+1/(D\sqrt{T})O(e^{-\delta^2 T}) .
\\
\end{aligned}
\end{equation}

Thus when we mutiply the above equation by $C$ we have
\begin{equation} C\cdot \hat{\phi}\to\frac{1}{\sqrt{2\pi(1-\sigma_2^2/\sigma_1^2)}} e^{-\frac{(\xi_1-\xi_2\sigma_2/\sigma_1)^2}{2(1-\sigma_2^2/\sigma_1^2)}}\end{equation}

as $T\to\infty.$

\textbf{Analysis of }$K_1:$\

Similar to previous proof,

we rewrite $$\hat{K}_1((n_1,t_1),x_1;(n_2,t_2),x_2)=\oint_{\Gamma_{1}} dw\oint_{\Gamma_{0}}dv\frac{e^{Tf_{0,1}(w)+\sqrt{T}f_{1,1}(w)}}{e^{Tf_{0,2}(v)+\sqrt{T}f_{1,2}(v)}}\frac{1}{w(w-v)}$$ where $f_{0,i}=u_i(w-\alpha\ln(w))+a_i(\ln(w-1)+\frac{\alpha}{1-\alpha}\ln(w)),$ and $f_{1,i}=\xi_i\sigma_i\ln(w).$

The derivatives of $f_{0,i}$ and $f_{1,i}$ are
\begin{equation}
\begin{aligned}
f'_{0,i}(w)&=\frac{(w-\alpha)(u_i(1-\alpha)(w-1)+a_i)}{w(w-1)(1-\alpha)},\\
f''_{0,i}(w)&=\frac{\alpha u_i}{w^2}-a_i(\frac{1}{(w-1)^2}+\frac{\alpha}{(1-\alpha)w^2}),\\
f''_{0,i}(\alpha)&=\frac{\sigma_i^2}{\alpha^2}>0,\\
f'_{1,i}(\alpha)&=\frac{\xi_1\sigma_i}{w}.\\
\end{aligned}
\end{equation}

There are two critical points for $f'_{0,i}(w)$, namely

\begin{equation}w_-=\alpha\ \text{and}\ w_{+,i}=1-\frac{a_i}{u_i(1-\alpha)}.\end{equation}

By using the inequality $n_i<(1-\alpha)^2 t$, $t_i=u_iT$ and $n_i=M+a_iT$, we have

$$M+a_it_i/u_)i<(1-\alpha )^2t_i$$
$$\Rightarrow a_i/u_i<(1-\alpha)^2<1$$
$$\Rightarrow 1-\frac{a_i/u_i}{1-\alpha}>\alpha>0$$
\begin{equation}\Rightarrow 0<w_-<w_{+,i}<1.\end{equation}

We choose $\Gamma_0=\{\alpha e^{i\phi},\phi\in[-\pi,\pi)\}$ and $\Gamma_1=\{v=1-re^{i\psi},r=1-w_{+2},\psi\in[-\pi,\pi)\}.$

Observing $|w-1|^2\geq (1-\alpha)^2\geq a_1/u_1$ we have
$$\frac{d\Re(f_{0,1}(w))}{d\phi}=-\frac{au_1\sin\phi}{|w-1|^2}(|w-1|^2-a_1/u_1),$$
which is decreasing away from $w=\alpha$ and from $|v|^2\geq |w_{+2}|\geq |w_{+2}|^2\alpha$ we have
$$\frac{d\Re(-f_{0,2}(v))}{d\psi}=-\frac{ru_2\sin\psi}{|v|^2}(|v|^2-\alpha(1-\frac{a_2}{u_2(1-\alpha})),$$
which is decreasing away from $v=w_{+2}.$

So now we have
$$\hat{K}_1((n_1,t_1),x_1;(n_2,t_2),x_2)$$
$$=\oint_{\Gamma_{1}} dw\oint_{\Gamma_{0}}dv \frac{e^{Tf_{0,1}(w)+\sqrt{T}f_{1,1}(w)}}{e^{Tf_{0,2}(v)+\sqrt{T}f_{1,2}(v)}}\frac{1}{w(w-v)}$$
$$=\frac{1}{C}\oint_{\Gamma_{1}} dw\oint_{\Gamma_{0}}dv \frac {\exp( T \Re\{ f_{0,1}(w_{-})-f_{0,1}(\alpha) \}+... )}{\exp( T \Re\{ f_{0,2}(w_{+2})-f_{0,2}(\alpha) \}+... )}\frac{1}{w(w-v)}$$
$$=\frac{1}{C}\oint_{\Gamma_{1}} dw\oint_{\Gamma_{0}}dv e^{-\delta T+...} \frac{1}{w(w-v)}$$
The leading term has an order of $e^{-\delta T}$ for some $\delta >0$ thus

$CK_1\to 0$ as $T\to\infty.$


\textbf{Analysis of }$K_2:$\

Again after substitution we have

$$\hat{K}_2((n_1,t_1),x_1;(n_2,t_w),x_2)=\oint_{\Gamma_{\alpha}}dv\oint_{\Gamma_{1,v}}dz   \oint_{\Gamma_{0}}\frac{dw}{w} \frac{(w-\alpha)^M}{(v-\alpha)^M}$$$$ \frac{e^{Tf_{0,1}(w)+\sqrt{T}f_{1,1}(w)}}{e^{Tf_{0,2}(z)+\sqrt{T}f_{1,2}(z)}}\frac{2(z-1)+A}{z-1+v-1+A}\frac{1}{z-v}\cdot$$
 $$\frac{(z-1+A)^M}{(v-1+A)^M}\frac{v-1+A}{(z-1+A)(w-\alpha)-(v-\alpha)(v-1+A)}. $$
\begin{equation}\end{equation}

Now we study the above formula in two cases:

\textbf{Pole at }$z=1$.\

The steepest descent path are
\\
$\{v=\alpha+\frac{\epsilon}{2} e^{i\psi},\psi\in [0,2\pi)\}$,
\\
$\{w=(\alpha-\epsilon)e^{i\phi},\phi\in [0,2\pi)\}$ since $w=\alpha$ is a root for the integrand, and also a saddle point,
\\
$\{z=1-re^{i\Omega}, r=1-w^{+}_2,\Omega\in [0,2\pi)\}.$
\\
For $\epsilon$ small enough the leading contribution is $\frac{1}{C}e^{-\delta T}$ for some $\delta>0.$ So as $T\to\infty$, the contribution from the pole at $z=1$ goes to zero.

\textbf{Pole at }$z=v$.\

Integrating the simple pole at $z=v$ we will have

$$\hat{K}_2((n_1,t_1),x_1;(n_2,t_w),x_2)=\oint_{\Gamma_{\alpha}}dv   \oint_{\Gamma_{0}}\frac{dw}{w} \frac{(w-\alpha)^M w^M}{(v-\alpha)^Mv^M}$$$$ \frac{e^{Tf_{0,1}(w)+\sqrt{T}f_{1,1}(w)}}{e^{Tf_{0,2}(v)+\sqrt{T}f_{1,2}(v)}} \frac{1}{w-v}. $$
\begin{equation}\end{equation}

Now we take the following steepest descent path $$\Gamma_\alpha=\{v=\alpha+R/\sqrt{T} e^{i\psi},\psi\in [0,2\pi)\}$$ and
$$\Gamma_{0}=\{w=(\alpha-L/\sqrt{T})e^{i\phi},\phi\in [0,2\pi),L>R\}.$$

Using Taylor series of

\begin{equation}
\begin{aligned}
f_{0,i}(w)&=f_{0,i}(\alpha)+\frac{(w-\alpha)^2}{2\alpha^2}\sigma_i^2+O((w-\alpha)^3),\\
f_{1,i}(w)&=f_{1,i}(\alpha)+\frac{\xi_i\sigma_i(w-\alpha)}{\alpha}+O((w-\alpha)^2),\\
(w(w-\alpha))^M&=\alpha^M(w-\alpha)^M+O((w-\alpha)^{M+1})
\end{aligned}
\end{equation}

and the change of variables $v-\alpha=\alpha V\sigma_1^{-1}T^{-1/2})$ and $w-\alpha=\alpha W\sigma_1^{-1}T^{-1/2}$ we have

\begin{equation}
\begin{aligned}
Tf_{0,1}(W)&=Tf_{0,1}(\alpha)+\frac{W^2}{2}+O(1/\sqrt{T}),\\
\sqrt{T}f_{1,1}(W)&=\sqrt{T}f_{1,1}(\alpha)+W\xi_1+O(1/\sqrt{T}),\\
Tf_{0,2}(V)&=Tf_{0,2}(\alpha)+\frac{V^2\sigma_2^2}{2\sigma_2^2}+O(1/\sqrt{T}),\\
\sqrt{T}f_{1,2}(V)&=f_{1,2}(\alpha)+V\xi_2\sigma_2/\sigma_1+O(1/\sqrt{T}),\\
\frac{(w(w-\alpha))^M}{(v(v-\alpha))^M}&=\frac{W^M}{V^M}\frac{1+O(\frac{1}{\sqrt{T}})}{1+O(\frac{1}{\sqrt{T}})}.
\end{aligned}
\end{equation}

Applying the above substitutions, we have

$$\hat{K}_2((n_1,t_1),x_1;(n_2,t_w),x_2)=\oint_{\Gamma_{\alpha}}dv   \oint_{\Gamma_{0}}\frac{dw}{w} \frac{(w-\alpha)^M w^M}{(v-\alpha)^Mv^M} \frac{e^{Tf_{0,1}(w)+\sqrt{T}f_{1,1}(w)}}{e^{Tf_{0,2}(v)+\sqrt{T}f_{1,2}(v)}} \frac{1}{w-v} $$

\begin{equation}=\frac{1}{D\sqrt{T\sigma^2_1}}\oint_{|V|=R}dV\oint_{-L+i\mathbb{R}}dW \frac{e^{W^2/2+W\xi_1}}{e^{V^2(\sigma_2/\sigma_1)^2/2+V\xi_2\sigma_2/\sigma_1}} \frac{W^M}{V^M}\frac{1}{W-V}+O(1/\sqrt{T}).\end{equation} Here $L>R.$

So now we have $$C\cdot\hat{K}_2((n_1,t_1),x_1;(n_2,t_w),x_2)\to\oint_{|V|=R}dV\oint_{-L+i\mathbb{R}}dW \frac{e^{W^2/2+W\xi_1}}{e^{V^2(\sigma_2/\sigma_1)^2/2+V\xi_2\sigma_2/\sigma_1}} \frac{W^M}{V^M}\frac{1}{W-V}$$
as $T\to\infty.$

$\square$
\newpage



\subsection{Linearly Decreasing Region}

\textbf{Theorem 4.2.1}\

For $n\in((1-\alpha)^2)t,t)$ or $\frac{\alpha ^2-2\alpha+2}{\alpha(2-\alpha)}<\pi(\theta)$,
The rescaled process  \begin{equation}X_T(\theta)=\frac{x_{[\nu t]}-(t-2\sqrt{\nu t^2- Mt })}{-T^{1/3}}\end{equation}
converges to the $\text{Airy}_2$

\begin{equation}\lim_{T\to\infty}X_T(\tau)=S_{v} A_2(\tau/S_h).\end{equation}.

\textbf{Proof:}

To prove this theorem, we have to show after rescaling our kernel $K=\hat{\phi}+K_1+K_2\to K_{Airy_{2}}.$

Similar to previous theorem, there are three parts to show

\begin{equation}
\begin{aligned}
&\lim_{T\to\infty} C \hat{\phi}= \sqrt{\frac{1}{4\pi(\frac{\tau_2}{S_h}-\frac{\tau_1}{S_h} )}}e^{-\frac{[(\frac{s_2}{S_v}-\frac{\tau_2^2}{S_h^2})-(\frac{s_1}{S_v}-\frac{\tau_1^2}{S_h^2})]^2}{4(\frac{\tau_2}{S_h}-\frac{\tau_1}{S_h})}}\mathbb{I}_{\{\tau_2>\tau_1\}},\\
&\lim_{T\to\infty} C K_1=\int_{\gamma_2}dV \int_{\gamma_1}dU \frac{e^{V^3/3+\tau_2S_h V^2- V (s_2/\mu-\kappa_1^2\tau_2^2/\kappa_0)/\kappa_0^{1/3}}}{e^{U^3/3+\tau_1S_h U^2-U ( s_1/\mu-\kappa_1^2\tau_1^2/\kappa_0)/\kappa_0^{1/3} }}\frac{1}{U-V} ,\\
&\lim_{T\to\infty} C K_2=0,
\end{aligned}
\end{equation} and the conjugation constant $C=\frac{e^{T^{2/3}f_{1,1}(w^*)+T^{1/3}f_{2,1}(w^*) }}{e^{T^{2/3}f_{1,2}(w^*)+T^{1/3}f_{2,2}(w^*) }}\kappa_0w^*T^{1/3}.$

First of all, our rescalings are
\begin{equation}t_i=T[\pi(\theta-\tau_iT^{-1/3})+\theta-\tau_iT^{-1/3}]\sim uT-\tau_i(1+\pi')T^{2/3}+\frac{1}{2}\pi'' \tau_i^2T^{2/3},\end{equation}
\begin{equation}n_i-M=\pi(\theta-\tau_iT^{-1/3})+\tau_i+T^{2/3}(1-\pi')\sim aT+\tau_i(1-\pi')T^{2/3}+\frac{1}{2}\pi'' \tau_i^2T^{2/3} ,\end{equation}

Using $\sqrt{1+\Delta x}=1+\frac{\Delta x}{2}-\frac{1}{8}\Delta^2 x+\cdots$,

$$x_i=t_i-2\sqrt{t_i(n_i-M)}-s_i(T^{1/3})$$$$
= u(1-2\sqrt{a/u})T-\tau_i T^{2/3}[(1+\pi')w^*+\frac{1-\pi'}{1-w^*}]
   +$$ \begin{equation}\tau_i^2T^{1/3}[\frac{1}{2}\pi''(\theta)(w^*-\frac{1}{1-w^*})+\frac{((1-\nu)\pi'-(1+\nu))^2}{4u(1-w^*)^3}]-s_i(T^{1/3}).\end{equation}

\textbf{Analysis of }$\hat{\phi}$:\

If we write $\hat{\phi}=\oint_{\Gamma_0}\frac{dw}{w} e^{g_1(w)T^{2/3}+g_2(w)T^{1/3}}\mathbb{I}_S,$ then assume $\tau_2>\tau_1$ for the rest of calculations
\begin{equation}
\begin{aligned}
g_1(w)&=(\tau_1-\tau_2)[-(1+\pi')w+(1-\pi') \ln((w-1)/w)+[(1+\pi')w^*+\frac{1-\pi'}{1-w^*}]\ln w]\\
      &=g_1(w^*)-(\tau_1-\tau_2)\frac{1}{2}(\frac{1+\pi'}{w^*}-\frac{\pi'-1}{w^*(1-w^*)^2})(w-w^*)^2+O((w-w^*))^3,\\
      &=g_1(w^*)-(\tau_1-\tau_2)\kappa_1 (w-w^*)^2+O((w-w^*))^3.	
\end{aligned}
\end{equation}

\begin{equation}
\begin{aligned}
g_2(w)&=[\frac{1}{2}\pi''(\theta)(w+\ln(\frac{w-1}{w}))-[\frac{1}{2}\pi''(\theta)(w^*-\frac{1}{1-w^*})+\frac{((1-\nu)\pi'-(1+\nu))^2}{4u(1-w^*)^3}]](\tau_1^2-\tau_2^2)\\&+(s_1-s_2)\ln w\\
      &=g_2(w^*)-[(\frac{\pi'-1}{2(1-w^*)^2w^*}-\frac{1+\pi'}{2 w^*})^2\frac{w^*(1-w^*)}{u}(\tau_1^2-\tau_2^2)-\frac{s_1-s_2}{w^*}](w-w^*)\\&+O((w-w^*)^2)\\
&=g_2(w^*)-(\frac{\kappa_1^2}{\kappa_0}(\tau_1^2-\tau_2^2)-\frac{s_1-s_2}{w^*})(w-w^*)+O((w-w^*)^2).
\end{aligned}
\end{equation}

So we have
\begin{equation}
\begin{aligned}
\hat{\phi}&=\oint_{\Gamma_0}\frac{dw}{w} e^{g_1(w)T^{2/3}+g_2(w)T^{1/3}}\\
      &=\oint_{\Gamma_0} e^{ T^{2/3}g_1(w^*)+ T^{1/3}g_2(w^*)-T^{2/3}(\tau_1-\tau_2)\kappa_1 (w-w^*)^2-T^{1/3}(\frac{\kappa_1^2}{\kappa_0}(\tau_1^2-\tau_2^2)-\frac{s_1-s_2}{w^*})(w-w^*) +T^{2/3}O((w-w^*))^3}.
\end{aligned}
\end{equation}

Let $(w-w^*)T^{1/3}\kappa_0^{1/3}=iy$ and using $\int_\mathbb{R} e^{-ax^2/2+iJx}dx=\sqrt{\frac{2\pi}{a}}e^{-J^2/(2a)}$, then it becomes
\begin{equation}\begin{aligned}
\hat{\phi}&= e^{ T^{2/3}g_1(w^*)+ T^{1/3}g_2(w^*)}\frac{1}{2\pi}T^{-1/3}\kappa_0^{-1/3}\int_{\mathbb{R}}e^{-(\tau_2-\tau_1)\kappa_1\kappa_0^{-2/3}y^2+iy(\kappa_1^2\kappa_0^{-4/3}(\tau_2^2-\tau_1^2)-\frac{s_2-s_1}{w^*\kappa_0^{1/3}})}dy+O(T^{-1/3})\\
&=e^{ T^{2/3}g_1(w^*)+ T^{1/3}g_2(w^*)}T^{-1/3}\kappa_0^{-1/3}{w^*}^{-1}\sqrt{\frac{1}{4\pi(\frac{\tau_2}{S_h}-\frac{\tau_1}{S_h} )}}e^{-\frac{[(\frac{s_2}{S_v}-\frac{\tau_2^2}{S_h^2})-(\frac{s_1}{S_v}-\frac{\tau_1^2}{S_h^2})]^2}{4(\frac{\tau_2}{S_h}-\frac{\tau_1}{S_h})}}+O(T^{-1/3})
\end{aligned}\end{equation}
where $S_h=\kappa_1^{-1}\kappa_0^{2/3}$ and $S_v=w^* \kappa_0^{1/3}.$

\textbf{Analysis of }$K_1$:\

\begin{equation}f_0(w)=wu+a\ln((w-1)/w)-[u(1-2\sqrt{a/u})+a]\ln w\end{equation}
\begin{equation}f_0'(w)=u(1+\frac{\nu}{w-1}-\frac{\nu}{w}-\frac{1-2\sqrt{\nu}}{w}) \end{equation}
One can check at $w^*=1-\sqrt{\nu}$ is a double pole such that $f_0'(w^*)=f_0''(w^*)=0.$
$$f_0(w)=f_0(w^*)-\frac{1}{3}\frac{u}{w^*(1-w^*)}(w-w^*)^3+O((w-w^*)^4).$$

\begin{equation}f_{1,i}(w)=-\tau_i(1+\pi')w+(1-\pi')\tau_i \ln((w-1)/w)+\tau_i[(1+\pi')w^*+\frac{1-\pi'}{1-w^*}]\ln w.\end{equation}

The Taylor series of the above functions are of the following:
\begin{equation}f_{1,i}(w)=f_{1,i}(w^*)-\tau_i\frac{1}{2}(\frac{1+\pi'}{w^*}-\frac{\pi'-1}{w^*(1-w^*)^2})(w-w^*)^2+O((w-w^*)^3)\end{equation}

\begin{equation}f_{2,i}(w)=[\frac{1}{2}\pi''(\theta)(w+\ln(\frac{w-1}{w}))-[\frac{1}{2}\pi''(\theta)(\alpha-\frac{1}{1-\alpha})+\frac{((1-\nu)\pi'-(1+\nu))^2}{4u(1-\alpha)^3}]]\tau_i^2\\+s_i\ln w\end{equation}

\begin{equation}f_2(w)=f_{2,i}(w^*)-[(\frac{\pi'-1}{2(1-w^*)^2w^*}-\frac{1+\pi'}{2 w^*})^2\frac{w^*(1-w^*)}{u}\tau_i^2-\frac{s_i}{w^*}](w-w^*)+O((w-w^*)^2)\end{equation}

If we let $\kappa_0=\frac{u}{w^*(1-w^*)}$ and $\kappa_1= \frac{1}{2}(\frac{1+\pi'}{w^*}-\frac{\pi'-1}{w^*(1-w^*)^2}),$

then

\begin{equation}
\begin{aligned}
f_0(w)&=f_0(w^*)-\frac{1}{3}\kappa_0(w-w^*)^3+O((w-w^*)^4),\\
f_{1,i}(w) &=f_{1,i}(w^*)-\tau_i\kappa_1(w-w^*)^2+O((w-w^*)^3),\\
f_{2,i}(w) &=f_{2,i}(w^*)-(\frac{\kappa_1^2}{\kappa_0}\tau_i^2-\frac{s_1}{w^*})(w-w^*)+O((w-w^*)^2).\\
.\end{aligned}
\end{equation}

With the rescalings, and substitute the above functions, we have

\begin{equation}K_1=B^{-1} \oint_{\Gamma_1}dv\oint_{\Gamma_0}dw\frac{e^{-T\frac{u}{3w^*(1-w^*)}(w-w^*)^3-T^{2/3}\tau_2\kappa_1(w-w^*)^2-T^{1/3}(\frac{\kappa_1^2\tau_2^2}{\kappa_0}-s_2/w^*)(w-w^*)}}{e^{-T\frac{u}{3w^*(1-w^*)}(v-w^*)^3-T^{2/3}\tau_1\kappa_1(v-w^*)^2-T^{1/3}(\frac{\kappa_1^2\tau_1^2}{\kappa_0}-s_1/\alpha)(v-w^*)}}\frac{1}{w(w-v)},\end{equation}

$B^{-1}=\frac{e^{T^{2/3}f_{1,2}(w^*)+T^{1/3}f_{2,2}(w^*) }}{e^{T^{2/3}f_{1,1}(w^*)+T^{1/3}f_{2,1}(w^*) }}.$

Let $(w-w^*)(\kappa_0T)^{1/3}=w_2$ and $(v-w^*)(\kappa_0T)^{1/3}=w_1$, then the above becomes
\begin{equation}1/C  \int_{\gamma_2}dw_2\int_{\gamma_1}dw_1\frac{e^{w_2^3+w_2^2\tau_2/S_h-(s_2/S_v-\tau_2^2/S_h^2)w_2}}{e^{w_1^3+\tau_1w_1^2/S_h-(s_1/S_v-\tau_1^2/S_h^2)w_1}}\frac{1}{w_1-w_2}+O(T^{-1/3}).\end{equation}


\textbf{Analysis of }$K_2$:\

After inserting the rescaling terms, we have

$$K_2=B^{-1} \oint_{\Gamma_1}dv\oint_{1,v} dz\oint_{\Gamma_0}dw\frac{e^{-T\frac{u}{3w^*(1-w^*)}(w-w^*)^3-T^{2/3}\tau_2\kappa_1(w-w^*)^2-T^{1/3}(\frac{\kappa_1^2\tau_2^2}{\kappa_0}-s_2/w^*)(w-w^*)}}{e^{-T\frac{u}{3w^*(1-w^*)}(z-w^*)^3-T^{2/3}\tau_1\kappa_1(z-w^*)^2-T^{1/3}(\frac{\kappa_1^2\tau_1^2}{\kappa_0}-s_1/\alpha)(z-w^*)}}\frac{1}{w(z-v)}$$

$$\frac{(w-\alpha)^M}{(v-\alpha)^M}\frac{2(z-1)+A}{z-1+v-1+A}\cdot\frac{(z-1+A)^M}{(v-1+A)^M}\frac{v-1+A}{(z-1+A)(w-\alpha)-(v-1+A)(v-\alpha)}$$
\begin{equation}\end{equation}
$B^{-1}=\frac{e^{T^{2/3}f_{1,2}(w^*)+T^{1/3}f_{2,2}(w^*) }}{e^{T^{2/3}f_{1,1}(w^*)+T^{1/3}f_{2,1}(w^*) }}.$

The leading comes from the $T^{-1/3}-$neighboorhood of $w^*,$
so we let $z=w^*+w_1(T\kappa_0)^{-1/3},\ w=w^*+w_2(T\kappa_0)^{-1/3},$ then

$$K_2= T^{-2/3}\kappa^{-2/3}B^{-1}\oint_{\Gamma_\alpha}dv\int_{\gamma_2}dw_2\int_{\gamma_1}dw_1\frac{e^{w_2^3/3+w_2^2\tau_2/S_h-(s_2/S_v-\tau_2^2/S_h^2)w_2}}{e^{w_1^3/3+w_1^2\tau_1/S_h-(s_1/S_v-\tau_1^2/S_h^2)w_1}} \times$$
\begin{equation}\frac{(w^*-\alpha)^M}{(v-\alpha)^M}\frac{2(w^*-1)+A}{w^*-1+v-1+A}\cdot\frac{(w^*-1+A)^M}{(v-1+A)^M}\frac{v-1+A}{(w^*-1+A)(w^*-\alpha)-(v-1+A)(v-\alpha)}.
\end{equation}

After multiplying by the conjugating constant, it goes to zero as $T$ goes to infinity.

\newpage
\subsection{Transition Process}
\textbf{Theorem 4.3.1}

When $n\sim (1-\alpha)^2t$, $\pi(\theta)=\frac{2-2\alpha+\alpha^2}{\alpha(2-\alpha)}\theta$ with $\theta$ fixed and we let the rescaled process be
\begin{equation}X_T(\tau)=\frac{x_n(t)-(t-2\sqrt{t(n-M)})}{-T^{1/3}}.\end{equation} We would like to show
\begin{equation}\lim_{T\to\infty}X_T(\tau)=S_v A_{DBM\to2}(\tau/S_h),\end{equation}where $S_h=\kappa_1^{-1}\kappa_0^{2/3}$ and $S_v=w^* \kappa_0^{1/3}.$


The process $A_{DBM\to2}$  also appeared in \cite{Bo2}.

\textbf{Definition:}

The $A_{DBM\to2}$ process is an $m-$point distributions at $\tau_1<\tau_2<...<\tau_m$ such that
\begin{equation}\mathbb{P}(\cap_{k=1}^m{A_{DBM\to2}(\tau_k)\leq s_k})=\det(\mathbb{I}-\chi_s K_{A_{DBM\to2}} \chi_s)_{L^2(\tau_1,...,\tau_m)\times \mathbb{R})}\end{equation}
where $\chi_s(\tau_k,x)=\mathbb{I}_{x>s_k}$ and
\begin{equation}K_{A_{DBM\to2}} =-\frac{-\frac{((s_2-\tau_2^2)-(s_1-\tau_1^2))^2}{4(\tau_2-\tau_1)}}{\sqrt{4\pi(\tau_2-\tau_1)}}\mathbb{I}_{\{\tau_2 >\tau_1\}}+\int_{\gamma_2}dw_2\int_{\gamma_1}dw_1\frac{e^{w_2^3+\tau_2w_2^2-(s_2-\tau_1^2)w_2}}{e^{w_1^3+\tau_1w_1^2-(s_1-\tau_1^2)w_1}}(\frac{w_1}{w_2})^M\frac{1}{w_1-w_2}. \end{equation}

Here we have $\gamma_1:e^{-2\pi/3}\infty\to e^{2\pi i/3}\infty$ and $\gamma_2: e^{\pi i/3}\infty\to e^{-\pi i/3}\infty$ provided that they both pass on the the left of $0$ and they don't meet each other.

\textbf{Proof:}\

To prove the convergence, we have to show after rescaling our kernel, $K=\hat{\phi}+K_1+K_2\to K_{A_{DBM\to 2}}.$

Similar to previous theorem, there are three parts to show
\begin{equation}
\begin{aligned}
&\lim_{T\to\infty} C \hat{\phi}= -\sqrt{\frac{1}{4\pi(\frac{\tau_2}{S_h}-\frac{\tau_1}{S_h} )}}e^{\frac{[(\frac{s_2}{S_v}-\frac{\tau_2^2}{S_h^2})-(\frac{s_1}{S_v}-\frac{\tau_1^2}{S_h^2})]^2}{4(\frac{\tau_2}{S_h}-\frac{\tau_1}{S_h})}}\mathbb{I}_{\tau_2>\tau_1},\\
&\lim_{T\to\infty} C K_1=\int_{\gamma_2}dw_2\int_{\gamma_1}dw_1\frac{e^{w_2^3/3+w_2^2\tau_2/S_v-(s_2/S_v-\tau_2^2/S_v^2)w_2}}{e^{w_1^3/3+w_1^2\tau_1/S_h-(s_1/S_v-\tau_1^2/S_h^2)w_1}}\frac{1}{w_1-w_2} ,\\
&\lim_{T\to\infty} C K_2=\oint_{\Gamma_0} du\int_{\gamma_2}dw_2\int_{\gamma_1}dw_1\frac{e^{w_2^3/3+w_2^2\tau_2/S_h-(s_2/S_v-\tau_2^2/S_h^2)w_2}}{e^{w_1^3/3+w_1^2\tau_1/S_h-(s_1/S_v-\tau_1^2/S_h^2)w_1}}\frac{1}{(w_1-u)(w_2-u)}\frac{w_1^M}{u^M}.
\end{aligned}
\end{equation}

Here $C=\frac{e^{\alpha(t_2-t_1)}(\alpha-1)^{n_2-n_1}}{\alpha^{x_2-x_1+n_2-n_1}}T^{1/3}$

From previous section, we have
\begin{equation}
\begin{aligned}
t_i&=T[\pi(\theta-\tau_iT^{-1/3})+\theta-\tau_iT^{-1/3}]\sim uT-\tau_i(1+\pi')T^{2/3}+\frac{1}{2}\pi'' \tau_i^2T^{2/3},\\
n_i-M &=\pi(\theta-\tau_iT^{-1/3})+\tau_i+T^{2/3}(1-\pi')\sim aT+\tau_i(1-\pi')T^{2/3}+\frac{1}{2}\pi'' \tau_i^2T^{2/3},\\
x_i &=u(1-2\sqrt{a/u})T-\tau_i T^{2/3}[(1+\pi')w^*+\frac{1-\pi'}{1-w^*}]\\
 &+\tau_i^2T^{1/3}\frac{((1-\nu)\pi'-(1+\nu))^2}{4u(1-w^*)^3}-s_i(T^{1/3})\\
 &=D_0 T-D_1\tau_iT^{2/3}+(D_2\tau_i^2-s_i)T^{1/3}.
\end{aligned}
\end{equation}

Also \begin{equation} w^*=1-\sqrt{\nu}=1-(1-\alpha)=\alpha. \end{equation}


If we write $\hat{\phi}=\oint_{\Gamma_0}\frac{dw}{w} e^{g_1(w)T^{2/3}+g_2(w)T^{1/3}}$\footnote{When $\pi(\theta)=\frac{2-2\alpha+\alpha^2}{\alpha(2-\alpha)}\theta$, $w^*=\alpha.$}then
\begin{equation}
\begin{aligned}
g_1(w)&=(\tau_1-\tau_2)[-(1+\pi')w+(1-\pi') \ln((w-1)/w)+[(1+\pi')\alpha+\frac{1-\pi'}{1-\alpha}]\ln w]\\
      &=g_1(\alpha)-(\tau_1-\tau_2)\frac{1}{2}(\frac{1+\pi'}{\alpha}-\frac{\pi'-1}{\alpha(1-\alpha)^2})(w-\alpha)^2+O((w-\alpha))^3,\\
      &=g_1(\alpha)-(\tau_1-\tau_2)\kappa_1 (w-\alpha)^2+O((w-\alpha))^3.	
\end{aligned}
\end{equation}

\begin{equation}
\begin{aligned}
g_2(w)&=[\frac{1}{2}\pi''(\theta)(w+\ln(\frac{w-1}{w}))-[\frac{1}{2}\pi''(\theta)(\alpha-\frac{1}{1-\alpha})+\frac{((1-\nu)\pi'-(1+\nu))^2}{4u(1-\alpha)^3}]](\tau_1^2-\tau_2^2)\\&+(s_1-s_2)\ln w\\
      &=g_2(\alpha)-[(\frac{\pi'-1}{2(1-\alpha)^2\alpha}-\frac{1+\pi'}{2 \alpha})^2\frac{\alpha(1-\alpha)}{u}(\tau_1^2-\tau_2^2)-\frac{s_1-s_2}{\alpha}](w-\alpha)+O((w-\alpha)^2)\\
&=g_2(\alpha)-(\frac{\kappa_1^2}{\kappa_0}(\tau_1^2-\tau_2^2)-\frac{s_1-s_2}{\alpha})(w-\alpha)+O((w-\alpha)^2).
\end{aligned}
\end{equation}

So we have
\begin{equation}
\begin{aligned}
\hat{\phi}&=\oint_{\Gamma_0}\frac{dw}{w} e^{g_1(w)T^{2/3}+g_2(w)T^{1/3}}\\
      &=\oint_{\Gamma_0} e^{ T^{2/3}g_1(\alpha)+ T^{1/3}g_2(\alpha)-T^{2/3}(\tau_1-\tau_2)\kappa_1 (w-\alpha)^2-T^{1/3}(\frac{\kappa_1^2}{\kappa_0}(\tau_1^2-\tau_2^2)-\frac{s_1-s_2}{\alpha})(w-\alpha) +T^{2/3}O((w-\alpha))^3}.
\end{aligned}
\end{equation}

Let $(w-\alpha)T^{1/3}\kappa_0^{1/3}=iy$ and using $\int_\mathbb{R} e^{-ax^2/2+iJx}dx=\sqrt{\frac{2\pi}{a}}e^{-J^2/(2a)}$, then it becomes
\begin{equation}\begin{aligned}
\hat{\phi}&= e^{ T^{2/3}g_1(\alpha)+ T^{1/3}g_2(\alpha)}\frac{1}{2\pi}T^{-1/3}\kappa_0^{-1/3}\int_{\mathbb{R}}e^{-(\tau_2-\tau_1)\kappa_1\kappa_0^{-2/3}y^2+iy(\kappa_1^2\kappa_0^{-4/3}(\tau_2^2-\tau_1^2)-\frac{s_2-s_1}{\alpha\kappa_0^{1/3}})}dy+O(T^{-1/3})\\
&=e^{ T^{2/3}g_1(\alpha)+ T^{1/3}g_2(\alpha)}T^{-1/3}\kappa_0^{-1/3}\alpha^{-1}\sqrt{\frac{1}{4\pi(\frac{\tau_2}{S_h}-\frac{\tau_1}{S_h} )}}e^{-\frac{[(\frac{s_2}{S_v}-\frac{\tau_2^2}{S_h^2})-(\frac{s_1}{S_v}-\frac{\tau_1^2}{S_h^2})]^2}{4(\frac{\tau_2}{S_h}-\frac{\tau_1}{S_h})}}+O(T^{-1/3})
\end{aligned}\end{equation}
where $S_h=\kappa_1^{-1}\kappa_0^{2/3}$ and $S_v=\alpha \kappa_0^{1/3}.$


Now we are going to show the rescaled kernel of $K_1$ converges to $$\frac{1}{(2\pi i)^2}\oint_{\Gamma_0}du\int_{\gamma_2}dw_2\int_{\gamma_1}dw_1\frac{e^{w_2^3+w_2^2\tau_2/S_h-(s_2/S_v-\tau_1^2/S_h^2)w_2}}{e^{w_1^3+\tau_1w_1^2/S_h-(s_1/S_v-\tau_1^2/S_h^2)w_1}}\frac{1}{w_1-w_2}.$$

We first write
\begin{equation}K_1=\oint_{\Gamma_1}dv\oint_{\Gamma_0}dw\frac{e^{Tf_0(w)+T^{2/3}f_{1,i}(w)+T^{1/3}f_{2,i}(w)+...}}{e^{Tf_0(v)+T^{2/3}f_{1,i}(v)+T^{1/3}f_{2,i}(v)+...}}\frac{1}{w(w-v)}\end{equation}

Similarly to previous section, the leading term is
\begin{equation}K_1=B^{-1} \oint_{\Gamma_1}dv\oint_{\Gamma_0}dw\frac{e^{-T\frac{u}{3\alpha(1-\alpha)}(w-\alpha)^3-T^{2/3}\tau_2\kappa_1(w-\alpha)^2-T^{1/3}(\frac{\kappa_1^2\tau_2^2}{\kappa_0}-s_2/\alpha)(w-\alpha)}}{e^{-T\frac{u}{3\alpha(1-\alpha)}(v-\alpha)^3-T^{2/3}\tau_1\kappa_1(v-\alpha)^2-T^{1/3}(\frac{\kappa_1^2\tau_1^2}{\kappa_0}-s_1/\alpha)(v-\alpha)}}\frac{1}{w(w-v)},\end{equation}

$B^{-1}=\frac{e^{T^{2/3}f_{1,2}(\alpha)+T^{1/3}f_{2,2}(\alpha) }}{e^{T^{2/3}f_{1,1}(\alpha)+T^{1/3}f_{2,1}(\alpha) }}.$

Let $(w-\alpha)(T\kappa_0)^{1/3}=w_2$ and $(v-\alpha)(T\kappa_0)^{1/3}=w_1$, then the above becomes
\begin{equation}1/C  \frac{1}{(2\pi i)^2}\int_{\gamma_2}dw_2\int_{\gamma_1}dw_1\frac{e^{w_2^3+w_2^2\tau_2/S_h-(s_2/S_v-\tau_2^2/S_h^2)w_2}}{e^{w_1^3+\tau_1w_1^2/S_h-(s_1/S_v-\tau_1^2/S_h^2)w_1}}\frac{1}{w_1-w_2}+O(T^{-1/3}),\end{equation}

$C^{-1}=B^{-1}\kappa_0^{-1} T^{-1/3}\alpha^{-1}$


Similarly, the rescaled kernel of $K_2$ converges to
$$\frac{1}{(2\pi i)^3}\oint_{\Gamma_0}du\int_{\gamma_2}dw_2\int_{\gamma_1}dw_1\frac{e^{w_2^3+w_2^2\tau_2/S_h-(s_2/S_v-\tau_2^2/S_h^2)w_2}}{e^{w_1^3+\tau_1w_1^2/S_h-(s_1/S_v-\tau_1^2/S_h^2)w_1}}\frac{1}{(w_1-u)(w_2-u)}(\frac{w_1}{u})^M$$

as $T\to\infty.$
\newpage

\section{Appendix}
\label{sec:LabelForAppendix:}

\subsection{Dyson's Brownian Motion}

In this section we define what a Dyson's Brownian Motion (DBM) is \cite{Me}.We introduced it from a random matrix
point of view \cite{Me}. Suppose, we have a matrix $H_{N\times N}$ with eigenvalues $x_j,1\leq j\leq N,$ then the
degree of freedom of the matrix is $d=N+N(N-1)\beta/2$ since the matrix $H$ is determined by its independent real
parameters which are $H_{ii},1\leq N;H^{(\lambda)}_{ij}, 1\leq i<j\leq N,0\leq\lambda\leq\beta-1.$ Now we rename other
parameters with $H_\mu$ with $\mu$ runs from $1$ to $d$ and use it to replace $\lambda, i,$ and $j$\footnote{$H_\mu$
are uncoupled and each is subjected to a fixed simple harmonic force. }. Suppose at time $t$, they have values
$H_1,..., H_d$ and $H_1+\delta H_1,...,H_d+\delta H_d$ at time $t+\delta t.$ Brownian motion is defined by saying each
infinitesimal $\delta H_\mu$ is a random variable such that

\begin{equation} <\delta H_\mu> f=-H_\mu \delta t\end{equation} \begin{equation} <(\delta H_\mu)^2> f=g_\mu kT\delta
t\end{equation} for some given constants $f,k\ \text{and}\ T.$ Here \begin{equation}
g_\mu=g^{(\lambda)_{ij}}=1+\delta_{ij}.\end{equation}

All other averages have a higher order in $\delta t.$ The Smoluchowski equation which corresponds to the above
requirements is

\begin{equation}f\frac{\partial P}{\partial t}=\sum_{\mu=1}^d (\frac{1}{2}g_\mu kT\frac{\partial^2}{\partial
H^2_\mu}+\frac{\partial}{\partial H_\mu}(H_\mu P)) \end{equation}

The solution $P(H_1,...,H_d;t)$ to the above equation is the time dependent joint probability density of $H_\mu.$ If
$H=H_1$ at time $t=\tau_1$, $H=H_2$ at time $t=\tau_1$,$f=1/2,1/\beta=2=kt$ and $H$ are $ M\times M$ hermitian
matrices ($N=M$) then we have the following formula\cite{Me}.

\textbf{Definition 5.1.1.}\

\begin{equation}P(\tau_1,H_1;\tau_2,H_2)=\frac{1}{(2\pi(1-e^{-2(\tau_2-\tau_1)}
))^M}\exp(-\frac{\text{Tr}(H_2-e^{\tau_1-\tau_2}H_1)^2}{2(1-e^{-2(\tau_2-\tau_1)})})\end{equation}

The joint distribution of the largest eigenvalue of the stationary DBM process are given by
\begin{equation}\mathbb{P}(\cap_{k=1}^m \text{DBM}(\tau_k)\leq
s_k)=\det(\mathbb{I}-\chi_sK^{\text{DBM}}\chi_s)_{L^2(\{\tau_1,...,\tau_m\}\times \mathbb{R})}\end{equation}

Here the kernel is given by \begin{equation} K^{{\rm DBM}}(\tau_1,x_1;\tau_2,x_2) \\ =-\frac{\exp\left(-{\displaystyle
\frac{(x_2-x_1e^{-(\tau_2-\tau_1)})^2}{2(1-e^{-2(\tau_2-\tau_1)})}}\right)}{\sqrt{2\pi(1-e^{-2(\tau_2-\tau_1)})}}\mathbb{I}_{[\tau_1<\tau_2]}
+  \sum_{k=1}^{M-1}e^{k(\tau_1-\tau_2)}p_k(x_1) p_k(x_2)e^{-x_2^2/2}\end{equation} where $p_k(x)=H_k(x/\sqrt{2})
\pi^{-1/4} 2^{-k/2} (k!)^{-1/2}$, and $H_k(x)$ is the standard Hermite polynomial of degree $k$ . $K_{\rm DBM}$ is
also called extended hermite kernel. \newpage


\subsection{Extended Hermite Kernel, Extended Airy Kernel and DBM to 2 Kernel}

In this section, we give an explanation about the double integral representation of the extended Hermite kernel of the
Dyson's Brownian motion $K_{DBM}$.

The extended Hermite kernel is defined as
\begin{equation}K_{DBM}(t,x;s,y)=\frac{1}{\pi(1-q^2)}\exp\{{-(\frac{(qx-y)^2}{1-q^2})}\}\chi_{t,s}+\sum_{k=0}^{n-1}ty
q^k p_k(x)p_k(y)e^{-y^2}\end{equation} where $p_k(x)$ are normalized Hermite polynomials,$\chi_{t,s}=\mathbb{I}_{t<s}$
and $q=e^{t-s}$. $p_k(x)$ can be written as $p_k(x)=\frac{1}{\sqrt{\sqrt{\pi}2^n n!}}H_n(x)$ \cite{AAR} where

\begin{equation}H_n(x)=\frac{n!}{2\pi i}\oint_\gamma  \frac{dz}{z^{n+1}}e^{-z^2-2zx},\end{equation}
\begin{equation}H_n(y)=\frac{2^n}{i\sqrt{\pi}}e^{y^2}\int_\Gamma dw e^{w^2-2wy}w^n.\end{equation}

Here the contour of $\Gamma= \{w\in-L+i\mathbb{R}\}$ and $\gamma=\{z=re^{i2\pi \phi},\phi\in[0,1), r<L\}$ with $L,r$
being some fixed positive constants.

Now we know $H_n(x)=0$ if $n\leq-1$, we can extend the following sum to $-\infty$, that is

$$\sum_{k=0}^{n-1}q^k p_k(x)p_k(y)e^{-y^2}=\sum_{k=-\infty}^\infty q^k p_k(x)p_k(y)e^{-y^2}$$
$$=\sum_{k=-\infty}^{n-1} \frac{2}{(2\pi i)^2}\int_{\Gamma}dw\oint_\gamma dz e^{w^2-2yw-z^2+2xz}\frac{q}{z}
\frac{q^kw^k}{z^k}.$$

After summing up the geometric series
$$\sum_{k=-\infty}^{n-1}\frac{q^kw^k}{z^k}=\frac{1}{w-z/q}\frac{q^nw^n}{z^{n-1}}$$ by requiring $|w/(zq)|<1$, we have

$$\sum_{k=0}^{n-1} q^k p_k(x)p_k(y)e^{-y^2}=\frac{2}{(2\pi i)^2}\int_{\Gamma}dw\oint_\gamma dz
e^{w^2-2yw-z^2+2xz}\frac{1}{w-z/q} \frac{q^nw^n}{z^{n}}$$

Now we use the change of variable $z\to zq$ then we have

\begin{equation}\sum_{k=0}^{n-1} q^k p_k(x)p_k(y)e^{-y^2}=\frac{2}{(2\pi i)^2}\int_{\Gamma}dw\oint_\gamma dz
e^{w^2-2yw-q^2z^2+2qxz}\frac{1}{w-z} \frac{w^n}{z^{n}}.\end{equation}

If we let $w\to -w/\sqrt{2}$ , $z\to -z/\sqrt{2}$ $x\to -x/\sqrt{2}$, and $y\to -y/\sqrt{2}$ we will have

\begin{equation}\sum_{k=0}^{n-1} q^k p_k(x)p_k(y)e^{-y^2}=\frac{1}{(2\pi i)^2}\oint_{|z|=r}dz \int_{L+i\mathbb{R}}dw
e^{w^2/2+yw-q^2z^2/2-zxq}\frac{1}{w-z} \frac{w^n}{z^{n}}.\end{equation}

Thus, an other way to express $K_{DBM}$ is
$$K_{DBM}(t,x;s,y)=\frac{1}{\pi(1-q^2)}\exp{-(\frac{(y-qx)^2}{2(1-q^2)})}\chi_{t,s}+$$ \begin{equation}\frac{1}{(2\pi
i)^2}\oint_{|z|=r}dz \int_{L+i\mathbb{R}}dw e^{w^2/2+yw-q^2z^2/2-zxq}\frac{1}{w-z} \frac{w^n}{z^{n}}.\end{equation}
which is what we had in previous section.

If we use \begin{equation}\frac{1}{\pi(1-q^2)}\exp\{{-(\frac{(qx-y)^2}{1-q^2})}\}\chi_{t,s}=\sum_{k=0}^\infty
p_k(x)p_k(t)q^ke^{-y^2},\end{equation} we also an alternative formula

\begin{equation} K_{DBM}(t,x;s,y)= \left\{\begin{array}{ll} \sum_{k=0}^{n-1} p_k(x)p_k(t)q^ke^{-y^2},& t\geq s,\\
-\sum_{k=n}^{\infty} p_k(x)p_k(t)q^ke^{-y^2}, & t<s. \end{array}\right. \end{equation}


Similarly we can also rewrite the extended Airy kernel

\begin{equation} K_{A_2}(\tau_1,s_1;\tau_2,s_2)= \left\{\begin{array}{ll} \int_0^\infty e^{-\lambda
(\tau_1-\tau_2)}\text{Ai}(s_1+\lambda)\text{Ai}(s_2+\lambda)d\lambda ,& \tau_1\geq \tau_2,\\ -\int^0_{-\infty}
e^{-\lambda (\tau_1-\tau_2)}\text{Ai}(s_1+\lambda)\text{Ai}(s_2+\lambda)d\lambda, & \tau_1<\tau_2. \end{array}\right.
\end{equation}
 as the following

$$e^{\frac{\tau_2^3-\tau_1^3}{3}-(\tau_2s_2-\tau_1s_1)}K_{A_2}(\tau_1,s_1;\tau_2,s_2)=-\frac{1}{\sqrt{4\pi(\tau_2-\tau_1)}}e^{\frac{[(s_2-\tau_2^2)-(s_1-\tau_1^2)]^2}{4(\tau_2-\tau_1)}}\mathbb{I}_{\{\tau_2>\tau_1\}}$$
\begin{equation}+\frac{1}{(2\pi
i)^2}\int_{\gamma_2}dw_2\int_{\gamma_1}dw_1\frac{e^{w_2^3+\tau_2w_2^2-(s_2-\tau_1^2)w_2}}{e^{w_1^3+\tau_1w_1^2-(s_1-\tau_1^2)w_1}}\frac{1}{w_1-w_2}
\end{equation}

by using \begin{equation}\frac{1}{w_1-w_2}=-\int_0^\infty e^{\lambda (w_1-w_2)} d\lambda,\end{equation}
\begin{equation}\frac{1}{\sqrt{4\pi(\tau_2-\tau_1)}}e^{-\frac{(s_2-s_1)^2}{4(\tau_2-\tau_1)}-\frac{(\tau_2-\tau_1)(s_1+s_2)}{2}+\frac{(\tau_2-\tau_1)^3}{12}}=\int_{-\infty}^\infty
e^{\lambda (\tau_2-\tau_1)}\text{Ai}(s_2+\lambda)\text{Ai}(s_1+\lambda) d\lambda\end{equation} from, \cite{Jo2} and
\begin{equation}\frac{-1}{2\pi i}\int_{\gamma_2}dw e^{w^3/3+aw^2+bw}=\text{Ai}(a^2-b)e^{2a^3/3-ab}.\end{equation}

Now we show how the kernel
$K_{DBM\to 2}$ in the proof of previous section can be written in the form of\footnote{$\bar{s_i}=s_i-\tau_i^2$.}

\begin{equation}-\frac{e^{-\frac{(\bar{s}_2-\bar{s}_1)^2}{4(\tau_2-\tau_1)}}}{\sqrt{4\pi(\tau_2-\tau_1)}}\mathbb{I}+\frac{1}{(2\pi
i)^2}\int_{\gamma_2}dw_2\int_{\gamma_1}dw_1
\frac{e^{w_2^3/3+\tau_2w_2^2-\bar{s}_2w_2}}{e^{w_1^3/3+\tau_1w_1^2-\bar{s}_1w_1}}\frac{w_1^M}{w_2^M}\frac{1}{w_1-w_2}.\end{equation}

The key is to use \begin{equation}\frac{1}{(w_1-u)(w_2-u)}=\frac{1}{w_1-w_2}(1/(w_2-u)-1/(w_1-u))\end{equation} and
$\frac{1}{w-u}=\frac{1}{w}\sum_{n=0}^\infty (u/w)^n, $ by letting $|u|<|w|.$

Once we use the identity and integrate out the integral on the $u$ plane, cancelation happens and we have $$
-\sqrt{\frac{1}{4\pi(\frac{\tau_2}{S_h}-\frac{\tau_1}{S_h}
)}}e^{\frac{[(\frac{s_2}{S_v}-\frac{\tau_2^2}{S_h^2})-(\frac{s_1}{S_v}-\frac{\tau_1^2}{S_h^2})]^2}{4(\frac{\tau_2}{S_h}-\frac{\tau_1}{S_h})}}\mathbb{I}_S
+\frac{1}{(2\pi
i)^2}\int_{\gamma_2}dw_2\int_{\gamma_1}dw_1\frac{e^{w_2^3+w_2^2\tau_2/S_v-(s_2/S_v-\tau_2^2/S_v^2)w_2}}{e^{w_1^3+w_1^2\tau_1/S_h-(s_1/S_v-\tau_1^2/S_h^2)w_1}}\frac{1}{w_1-w_2}
$$ $$+\frac{1}{(2\pi i)^3}\oint_{\Gamma_0}
du\int_{\gamma_2}dw_2\int_{\gamma_1}dw_1\frac{e^{w_2^3+w_2^2\tau_2/S_h-(s_2/S_v-\tau_2^2/S_h^2)w_2}}{e^{w_1^3+w_1^2\tau_1/S_h-(s_1/S_v-\tau_1^2/S_h^2)w_1}}\frac{1}{(w_1-u)(w_2-u)}\frac{w_1^M}{u^M}
=$$

$$-\sqrt{\frac{1}{4\pi(\frac{\tau_2}{S_h}-\frac{\tau_1}{S_h}
)}}e^{\frac{[(\frac{s_2}{S_v}-\frac{\tau_2^2}{S_h^2})-(\frac{s_1}{S_v}-\frac{\tau_1^2}{S_h^2})]^2}{4(\frac{\tau_2}{S_h}-\frac{\tau_1}{S_h})}}\mathbb{I}_{\tau_2>\tau_1}+$$$$
+\frac{1}{(2\pi
i)^2}\int_{\gamma_2}dw_2\int_{\gamma_1}dw_1\frac{e^{w_2^3+w_2^2\tau_2/S_h-(s_2/S_v-\tau_2^2/S_h^2)w_2}}{e^{w_1^3+w_1^2\tau_1/S_h-(s_1/S_v-\tau_1^2/S_h^2)w_1}}\frac{w_1^M}{w_2^M}\frac{1}{w_1-w_2}.$$
\begin{equation}\end{equation}

\textbf{Acknowledgement}\

 The author offers his sincerest thanks to Professor Craig Tracy, who has supported him throughout this paper with his long hours of patient guidance whilst allowing me the room to work in my own way. This work was supported by National Science Foundation through the grant number DSM0906387.

\end{document}